\documentclass[nonatbib]{elsarticle/elsarticle}

\usepackage{hyperref}
\hypersetup{allcolors=blue, colorlinks=true}
\usepackage[utf8]{inputenc}

\newcommand{\email}[1]{\texttt{#1}}

\author[1]{George G. Vega Yon\corref{cor1}}
\ead{vegayon@usc.edu}
\author[2]{Andrew Slaughter}
\ead{andrew.j.slaughter.civ@mail.mil}
\author[1]{Kayla de la Haye}
\ead{delahaye@usc.edu}

\cortext[cor1]{Corresponding author}
\address[1]{Department of Preventive Medicine, University of Southern California, USA}
\address[2]{US Army Research Institute for the Behavioral and Social Sciences, USA}

\usepackage{amsmath,amssymb}
\usepackage{tabularx,booktabs}
\usepackage{graphicx,multicol}

\newcommand{\isone}[1]{{\boldsymbol{1}\left( #1 \right)}}

\newcommand{\Prcond}[2]{{\mbox{Pr}\left(#1\vphantom{#2}\;\right|\left.\vphantom{#1}#2\right)}}

\newcommand{\sufstats}[1]{s\left(#1\right)}
\renewcommand{\exp}[1]{\mbox{exp}\left\{#1\right\}}
\newcommand{\transpose}[1]{{#1}^\mathbf{t}}

\newcommand{\params}{\theta}

\newcommand{\Adjmat}{Y}
\newcommand{\adjmat}{y}
\newcommand{\ADJMAT}{\mathcal{Y}}

\newcommand{\Indepvar}{X}

\newcommand{\normconst}{\kappa\left(\params, \Indepvar\right)}

\graphicspath{{./figures/}{.}{./terms/}}

\usepackage{pstricks}

\definecolor{USCCardinal}{HTML}{990000} % 153 0 0 in RGB
\definecolor{USCGold}{HTML}{FFCC00}
\definecolor{USCGray}{HTML}{CCCCCC}

\def\ergmito{ERGM\textit{ito}}
\def\ergmitos{\ergmito{}\textit{s}}

\makeatletter
\let\c@author\relax
\makeatother

\usepackage[style=numeric-comp]{biblatex}
\usepackage[normalem]{ulem} 

\usepackage{lineno}
\usepackage{textcomp} % FOR USING THE COPYRIGHT SYMBOL

\title{Exponential Random Graph models for Little Networks\tnoteref{t1}}

\tnotetext[t1]{The authors would like to thank Garry Robins, Carter Butts, Johan Koskinen, and Noshir Contractor for their valuable contributions to this work. Authors Contact information: \email{vegayon@usc.edu}, \email{andrew.j.slaughter.civ@mail.mil}, \email{delahaye@usc.edu}. \textcopyright{} 2019. This work is licensed under a Creative Commons Attribution-NonCommercial-NoDerivatives 4.0 International License.}

\date{May, 2020}

\hypersetup{pdfauthor=The Avengers and Fantastic four}

\begin{document}

\begin{abstract}
    Statistical models for social networks have enabled researchers to study complex social phenomena that give rise to observed patterns of relationships among social actors and to gain a rich understanding of the interdependent nature of social ties and actors. Much of this research has focused on social networks within medium to large social groups. To date, these advances in statistical models for social networks, and in particular, of Exponential-Family Random Graph Models (ERGMS), have rarely been applied to the study of small networks, despite small network data in teams, families, and personal networks being common in many fields. In this paper, we revisit the estimation of ERGMs for small networks and propose using exhaustive enumeration when possible. We developed an R package that implements the estimation of pooled ERGMs for small networks using Maximum Likelihood Estimation (MLE), called ``ergmito''. Based on the results of an extensive simulation study to assess the properties of the MLE estimator, we conclude that there are several benefits of direct MLE estimation compared to approximate methods and that this creates opportunities for valuable methodological innovations that can be applied to modeling social networks with ERGMs.
\end{abstract}


\maketitle

\section{Acknowledgement}

 This material is based upon work support by, or in part by, the U.S. Army Research Laboratory and the U.S. Army Research Office under grant number W911NF-15-1-0577. The views, opinions, and/or findings contained in this paper are those of the authors and shall not be construed as an official Department of the Army position, policy, or decision, unless so designated by other documents.
 
 Computation for the work described in this paper was generously supported by the University of Southern California’s Center for High-Performance Computing (hpcc.usc.edu).

% \clearpage
\section{Introduction}

Statistical models for social networks have enabled researchers to study complex social phenomena that give rise to observed patterns of relationships among social actors, and to gain a rich understanding of the \textit{interdependent} nature of social ties and social actors \cite{Snijders2011,lusher2012exponential}. For example, this research has provided new insights into the role that the attributes of social actors (e.g., their characteristics, beliefs, and decisions), and endogenous structural processes (e.g., social balance, and relationship reciprocity) play in shaping social networks across different populations and social settings, and how  these social networks, in turn, influence individuals and groups. 

Much of this research has focused on social networks within medium to large social groups: networks ranging from dozens or hundreds of members (e.g., classrooms and organizations) to millions (e.g., online social networks). However, modern advances in statistical models for social networks have rarely been applied to the study of small networks, despite small network data from teams, families, and personal (ego-centric) networks being common in many fields that study social phenomena \cite{HENTTONEN201074, CarterDR2015, bott2001family,crossley2015social}. The study of small networks often uses descriptive statistics that summarize basic structural features of the network; for example, the density, degree distribution, or triad count. However, researchers in these fields are often interested in testing hypotheses about \textit{why} localized social structures, such as reciprocity, balance, and homophily, emerge in these small groups. A key limitation to such work has been the availability of statistical models for networks that can flexibly test and control for the kind of dependencies inherent to network data. In this paper, we propose an approach for applying one of the most widely used statistical models for social networks--exponential random graph models, or ERGMs--to small graphs, to enable new research on ``little networks''. 

\section{Exponential-Family Random Graph Models}

Exponential-family random graph models (ERGMs) are one of the most popular tools used by social scientists to understand social networks and test hypotheses about these networks  \cite[][and others]{Robins2007,Holland1981,Frank1986,Wasserman1996,Snijders2006}. In this family of models, an observed graph $\adjmat$ , comprised of a set of nodes (vertices) and ties (edges), is characterized by a set of sufficient statistics defined on the graph, $\sufstats{\adjmat}$, and parameters $\params$. In a model that also includes node characteristics $\Indepvar$, this leads to the following equation:

\begin{equation}
\label{eq:ergm}
  \Prcond{\Adjmat = \adjmat}{\params, \Indepvar} = \frac{%
  	\exp{\transpose{\params}\sufstats{\adjmat, \Indepvar}}%	
  }{
  	\kappa\left(\params, \Indepvar\right)
  },\quad\forall \adjmat\in\ADJMAT
\end{equation}

\noindent Where $\normconst{} = \sum_{\adjmat\in\ADJMAT}\exp{\transpose{\theta}\sufstats{\adjmat, \Indepvar}}$ is the normalizing constant, and $\ADJMAT$ is the support of the model that is usually assumed to include all graphs of the same type (e.g., directed or undirected) and size, that do not include self-ties. In the directed graph case, the size of $\ADJMAT$ equals $2^{n(n-1)}$ possible graphs. This makes the exact calculation of $\normconst{}$, and therefore of \eqref{eq:ergm},  computationally expensive. A sophisticated array of parameters can be specified for ERGMs that reflect social and structural process of interest to social scientists, such as social closure, connectivity, and other affiliation preferences.  \autoref{fig:ergm-structs} shows some examples of the structures (statistics) that can be estimated with ERGMs.

\def\fig1width{.45\linewidth}
\begin{figure*}[tb]
\centering
\begin{tabular}{m{.2\linewidth}<\centering m{.4\linewidth}<\raggedright}
\toprule Representation & Description  \\ \midrule
\includegraphics[width=\fig1width]{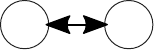} & Mutual Ties (Reciprocity)\linebreak[4]$\sum_{i\neq j}y_{ij}y_{ji}$  \\
\includegraphics[width=\fig1width]{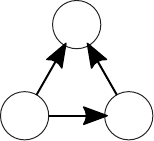} & Transitive Triad (Balance)\linebreak[4]$\sum_{i\neq j\neq k}y_{ij}y_{jk}y_{ik}$  \\
\includegraphics[width=\fig1width]{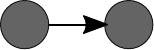} & Homophily\linebreak[4]$\sum_{i\neq j}y_{ij}\mathbf{1}\left(x_i=x_j\right)$ \\
\includegraphics[width=\fig1width]{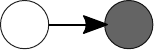} & Attribute-receiver effect \linebreak[4]$\sum_{i\neq j}y_{ij}x_j$ \\
\includegraphics[width=\fig1width]{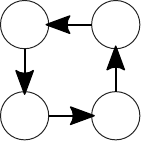} & Four Cycle\linebreak[4]$\sum_{i\neq j \neq k \neq l}y_{ij}y_{jk}y_{kl}y_{li}$  \\
\bottomrule
\end{tabular}
\caption{\label{fig:ergm-structs}Besides of the common edge count statistic (number of ties in a graph), ERGMs allow measuring other more complex structures that can be captured as sufficient statistics. }
\end{figure*}

While other methods for studying small graphs exist, e.g. non-parametric tests like the Conditionally Uniform Graph tests (CUG tests in the \textit{social networks} literature \cite{Anderson1999,Faust2010} and rewiring algorithms in the \textit{network science} literature \cite{Milo2004a,Milo2004b}), all in all, ERGMs have more flexibility because they can be used to test complex hypotheses in a multivariate framework. As noted in \cite{Butts2008}, most of these non-parametric methods can be written in the form of \eqref{eq:ergm}, which means that ERGMs can be viewed as a generalized version of many of these tests.

Although ``small networks'' is a topic mentioned several times in the literature on social network models  \cite{Wasserman1996,Frank1986,Snijders2011},  interest in larger social networks has dominated the field.\footnote{This is perhaps because, as put by \cite{Snijders2011}, small networks are considered to be ``uninteresting special cases''} Thus, ERGM methods have been developed to accommodate larger networks (although it is only very recent developments that have begun to scale well to ``very large'' networks of several thousand nodes or more \cite{Stivala2020}). One example of this is the calculation of the likelihood function: rather than being calculated using exhaustive enumeration (which we will refer to as "exact likelihood"), the most popular software packages used for estimating these models apply simulation-based estimation methods. As a consequence, current methods used to estimate ERGMs for medium to large networks do not translate well to small network data (i.e., 6 or fewer nodes in a directed network), and applications of these statistical network models to small networks are rare.  

One major technical and theoretical issue in ERGM estimation generally, which is exacerbated with small networks, is the problem of \textit{non-existence of Maximum Likelihood Estimation (MLE)}. Non-existence of MLEs (or the convex-hull problem) occurs when the observed graph's statistics lie in a region on or near the boundary of the support \cite{Barndorff-Nielsen2014}, and can be stressed when estimation depends on Monte Carlo Integration \cite{Handcock2003}. Small networks, which are more likely to be nearly empty or nearly full, have a smaller region of support, and are more likely to be on or near the boundary of that support. For example, if we are trying to estimate an ERGM in a network with only three nodes, in the scenario where the graph is directed and does not allow for self-ties, the chances of obtaining a graph with either one or zero ties (i.e., empty or almost completely empty), or a graph with five or six ties (i.e., fully or almost fully connected) is about 20\% using a uniform sampler.\footnote{For more on the discussion on existence, degeneracy, and instability see \cite{Jacobsen1989,Rinaldo2009,Schweinberger2011}.}

Because researchers studying small networks often have observed \textit{samples} of small networks (e.g., multiple team, family, or personal/egocentric networks), a common work-around to the issue of non-existence of MLE is to combine the independent small networks into a single larger block-diagonal graph. Estimation then proceeds by assuming that ties between blocks are impossible (i.e., treated as structural zeros in estimation). The major problems with this approach are that it can be complicated to fit, and difficult to extend. As an example of the former, the same set of constraints (the structural zeros) that allow for the model to be fit can also make the estimation procedure more difficult, and increase the possibility of sampling problems during MCMC estimation. However, a more important challenge with the block-diagonal approach are difficulties with extension. A basic ``complete pooling'' model, which assumes a common data generating process across all networks, is straightforward to define. However, relaxing that assumption to allow for variability across graphs (i.e., unpooled or partially-pooled models) can be problematic; it would typically require the creation of block-wise node membership attributes, and complex interaction terms involving subgraph statistics and node membership variables. Moreover, extending this framework to not only  allow for between-group variability, but to explicitly  \textit{predict} it (for example, as a function of additional group-level variables), is not straightforward with this complete-pooling approach.

To overcome the challenges described above for fitting ERGMs to small networks, we leverage the fact that in the case of small networks, the full likelihood function \textit{is} tractable. This allows the direct estimation of model parameters without using Markov Chain Monte Carlo (MCMC) or other approximate methods, avoiding some of the convergence issues associated with the convex-hull problem \cite{Handcock2003}. It also makes it much easier to combine ERGMs with other statistical techniques, opening the door for many possibilities of richer methods to model and understand small-group network structure and dynamics. In this paper, we describe how modern computational power allows for the complete specification of the likelihood for small graphs, and how this specification allows us to use the standard tools of MLE, instead of approximate methods. We present examples using these techniques; provide some initial results on empirical bias, type I error rates, and power based on a simulation study; illustrate the flexibility of this method with an empirical application; and discuss future extensions these techniques make feasible.

\section{\ergmitos{}: ERGMs for small networks}

With modern computers, calculating the exact likelihood function of an ERGM for a small network becomes computationally feasible. This has an important implication: the process for estimating the parameters of an ERGM for small networks can be done directly. Many innovative techniques have been developed to handle models with intractable normalizing constants (e.g., Markov Chain Monte Carlo [MCMC] based estimation methods, Bayesian techniques such as the exchange sampler, etc.), and often these techniques work quite well. Of course, no techniques are without tradeoffs; MCMC-based estimation can be sensitive to starting values, and the quality of standard errors can depend on the availability of analytic gradients \cite{Park2018}. Bayesian techniques like the exchange sampler \cite{Moller2006} may be comparatively slow, which may be an issue when many networks are to be analyzed.

Moreover, simulation-based methods may have particular susceptibilities to the convex-hull problem. As stated by \cite[p. 7]{Handcock2003}, ``[i]f the model used to simulate the graphs is not close enough to produce realizations that cover the observed values of the statistics, the MC-MLE will not exist even in cases where the MLE does.'' For example, many common network models, such as triangle-based models, can lead to bimodal distributions of graph statistics that simulation-based methods have difficulty with; even when the MLE falls between the modes \cite{Hunteretal2012}. Therefore, even though the non-existence issue is not completely avoided, a method based on exact (non-simulation) inference may not only provide a better solution (in general) by avoiding the additional uncertainty induced by simulations and approximations, but it may also help to mitigate the problem in cases where the MLE exists.

Of course, the statistical analysis of a single small network could be uninformative due to the small numbers of dyads, and a high restriction in the variability of possible subgraph statistics. Fortunately, research on small networks typically involves collecting data from \textit{samples} of small groups (vs. the more typical 'case studies' of single larger networks), which allows for the development of models to analyze structural variation both within and across small networks. If we assume that the sample of networks comes from a population of networks (groups) that are governed by the same data generating process, we end up with the following likelihood, defining a completely-pooled model:

\begin{multline}
    \label{eq:ergm-pooled}
    \Prcond{\Adjmat_1 = \adjmat_1, \dots, \Adjmat_P = \adjmat_P}{\params, \Indepvar_1, \dots, \Indepvar_p} = \\
    \prod_{p=1}^P\frac{%
    		\exp{\transpose{\params}\sufstats{\adjmat_p, \Indepvar_p}}%	
    	}{
    		\kappa_p\left(\params, \Indepvar_p\right)
    	}
\end{multline}

\noindent Where $P$ denotes the number of networks used in the model, and $\kappa_p\left(\params, \Indepvar_p\right)$ is explicitly calculated, unlike existing approaches to ERGM estimation. We call this framework, which is a revisited version of ERGM in the case of small networks, \ergmito{}. In general, this extension can be feasibly applied to small graphs containing at most 6 nodes if directed, or 8 if undirected. %\footnote{The \textit{ito}/\textit{ita} suffix is used in Spanish to denote small, or affection. We are especially grateful to George Barnett who proposed the name during the North American Social Networks Conference in 2018.}.

Not to be confused with \textit{pooled estimators} -- i.e. aggregating various parameter estimates from independent model fits-- pooled-data models have several benefits, including the ability to consider small networks that otherwise would be excluded from an analysis; e.g., because they are fully connected or empty graphs. Moreover, as we will emphasize later in \autoref{sec:simulation-study}, as long as at least one network in the sample has values on the boundary for each type of sufficient statistic, the MLEs will generally exist \cite[see][]{Handcock2003}.

One issue that may be of concern is the feasibility of the underlying assumptions when estimating pooled-data models with networks of different sizes. Because parameter estimates often encode network size, one may argue that pooling networks of different sizes into a single model may not be appropriate. However, there are several ways to control for size-induced heterogeneity; for example, including fixed or random effects at the graph level to account for size, or using approaches such as those described in \cite{Krivitsky2011,Krivitsky2015,Butts2015}. In the cases presented in this paper, we focus on samples of networks that are of similar sizes (networks of size 4 and 5); thus, these issues are unlikely to be of great concern within a small range of values, although we demonstrate how they can be accounted for in our applied example (\autoref{sec:empirical}).

% This means caution is required when combining or comparing parameter estimates from different networks. Even the interpretation of the basic edge count parameter can become difficult; as pointed out in \cite{Krivitsky2011}, assuming a common edge parameter in networks of different sizes is equivalent to assuming equal edge probabilities in an edge-only model. However, many real-world social networks are sparse, with a density that grows much more slowly than network size. 

In the following sections we illustrate and investigate the properties of estimating ERGMs for small networks using this approach. All simulations and model fitting were conducted using the R package \textit{ergmito}, which has been developed to implement the methods described in this paper.

\section{Illustration with simulated data: fivenets}

\subsection{Data-generating-process and model fitting}

\begin{figure}[tb]
    \centering
    \includegraphics[width=.6\linewidth]{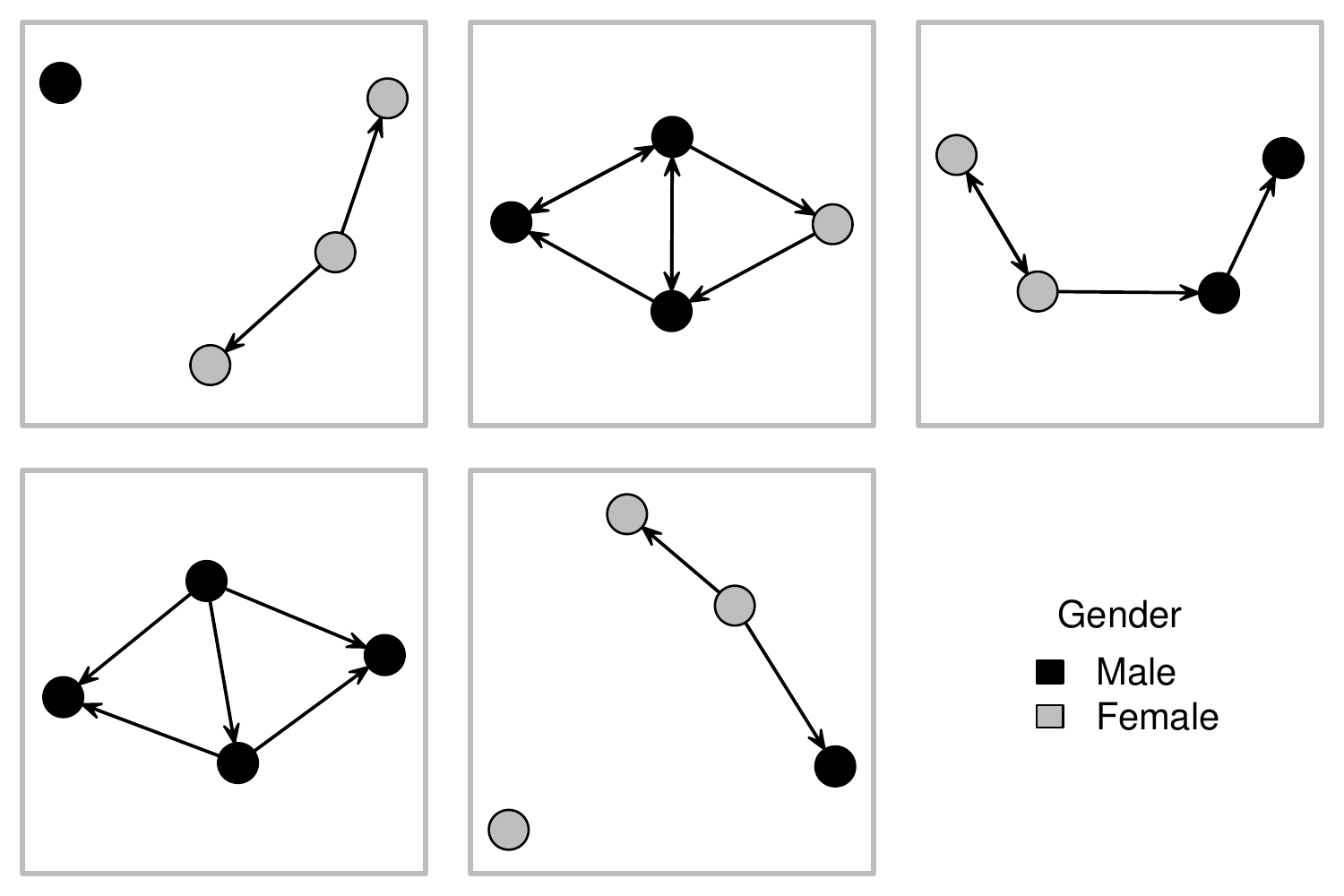}
    \caption{\label{fig:fivenets}Fivenets data set. These graphs were randomly drawn from an ERGM distribution with two parameters: number of edges and gender homophily, with parameters equal to -2.0 and 2.0 respectively.}
\end{figure}

Starting with a s simple example, we now look at a simulated data set that was created using the data-generating-process of \ergmitos{}. This particular dataset, which we call ``fivenets'', is included in the in the R package \textit{ergmito}\footnote{The R package is available to be downloaded at \url{https://github.com/muriteams/ergmito}.}.
The data set contains five small graphs with nodal attributes (we use gender in the following example), with the networks generated using the following specification:

\begin{multline*}
\Prcond{\Adjmat = \adjmat}{\Indepvar, \params} = \\
\frac{ %
    \exp{\params_{edges}\left(\sum_{i,j} \adjmat_{ij}\right) + %
    \params_{same}\left(\sum_{i,j} \adjmat_{ij}\isone{\Indepvar_{i} = \Indepvar_{j}}\right)} %
    }{%
    \normconst{}
    }
\end{multline*}

\noindent where $\params_{edges} = -2.0$ and $\params_{same} = 2.0$. Using this equation we draw five networks of size four. The process of ``homophily'' is represented by a parameter that is defined as the number of ties in which ego and alter have the same gender, $\params_{same}$. Before drawing the networks we randomly generated the node attribute (gender) to each vertex as a Bernoulli with parameter 0.5. \autoref{fig:fivenets} shows the generated networks, including their nodal attributes.

Using the \textit{ergmito} R package, we fit three different models to the data: (1) a Bernoulli graph, which is a model that only includes the ``edges'' parameter, (2) a model with ``gender homophily'' as its only parameter, and finally (3) a model including both ``edges'' and ``gender homophily'', which is the correct specification of the model. Some details regarding the computational aspects of the model fitting process are provided in \ref{appendix:mle}.

In general, while practitioners are accustomed to dealing with a single set of observed sufficient statistics, sometimes called ``target'' statistics, pooled models instead feature an array of such statistics. \autoref{tab:obs-suff-fivenets} displays the counts used in this model, from the Fivenets data.

% latex table generated in R 3.6.3 by xtable 1.8-4 package
% Wed Apr 29 09:52:45 2020
\begin{table}[ht]
\centering
\begin{tabular}{l*{2}{m{.2\linewidth}<\centering}}
  \toprule
Net id & edgecount & count of gender homophilic ties \\ 
  \midrule
  1 &   2 &   2 \\ 
    2 &   7 &   5 \\ 
    3 &   4 &   3 \\ 
    4 &   5 &   5 \\ 
    5 &   2 &   1 \\ 
   \bottomrule
\end{tabular}
\caption{\label{tab:obs-suff-fivenets}Observed sufficient for the \textit{fivenets} dataset. In the case of pooled-data models, there is no one set of observed (target) sufficient statistics, but an array of such statistics. This table shows the \textit{edgecount} and the \textit{count of gender homophilic ties} in the \textit{fivenets} dataset.}
\end{table}

\autoref{table:coefficients} shows the estimation results of the three different specifications of the model and, as expected, model (3) has the best overall fit to the data. Furthermore, since all three models were fitted using MLE, we can compare the edgecount and homophily models with the full model using Likelihood Ratio tests \cite{Zeileis2002}.

\begin{table*}[ht]
\begin{center}
\begin{tabular}{l c c c}
\toprule
 & Homopholy & Edgecount & Full model \\
\midrule
Edgecount                    &              & $-0.69^{*}$    & $-1.70^{**}$ \\
                             &              & $(0.27)$       & $(0.54)$     \\
Homophily (on Gender)        & $-0.12$      &                & $1.59^{*}$   \\
                             & $(0.34)$     &                & $(0.64)$     \\
\midrule
LR-test statistic ($\chi^2$) & $7.04 ^{**}$ & $13.72 ^{***}$ & $$           \\
AIC                          & $85.06$      & $78.38$        & $73.34$      \\
BIC                          & $87.15$      & $80.48$        & $77.53$      \\
Log Likelihood               & $-41.53$     & $-38.19$       & $-34.67$     \\
Num. networks                & $5$          & $5$            & $5$          \\
% Convergence                  & $0$          & $0$            & $0$          \\
\bottomrule
\multicolumn{4}{l}{\scriptsize{$^{***}p<0.001$; $^{**}p<0.01$; $^{*}p<0.05$}}
\end{tabular}
\caption{Fitted ERGMitos using the fivenets dataset. Looking at AICs and LR-test statistics, the full model (last column of the table) is the one with the best fit to the observed data. More over, the 95\% level CI of each covers the true parameters: $\hat\theta_{edges} \in [-2.77, -0.64]$; $\hat\theta_{Homophily} \in [0.33, 2.85]$.}
\label{table:coefficients}
\end{center}
\end{table*}

It is important to note that the \textit{ergm} package can also be used to calculate exact likelihoods, and that this feature has been available for a long time. Some of the additional features and extensions provided in the \textit{ergmito} package, which are illustrated in subsequent sections of the paper, are: a simple way of estimating pooled-data models, simulating small networks using exact likelihoods, evaluating goodness-of-fit at the graph level for pooled-data models, and including arbitrary effects like interaction effects and transformation of the canonical ERGM terms. The \textit{goodness of fit} of this model is evaluated in the following section.

\subsection{Goodness-of-fit in \ergmitos}

\begin{figure}[tb]
    \centering
    \caption{Goodness-of-fit in \ergmitos{}. This illustrates how the observed sufficient statistics of each one of the 5 networks (x-axis) locate in the overall estimated distribution based on the fitted \ergmito{}. The gray lines in each box show the minimum and maximum value that the sufficient statistics can take in each one of the 5 networks, whereas the dotted lines provide a 90\% confidence interval. The dots are the observed statistics in each network.}
    \includegraphics[width=.7\linewidth]{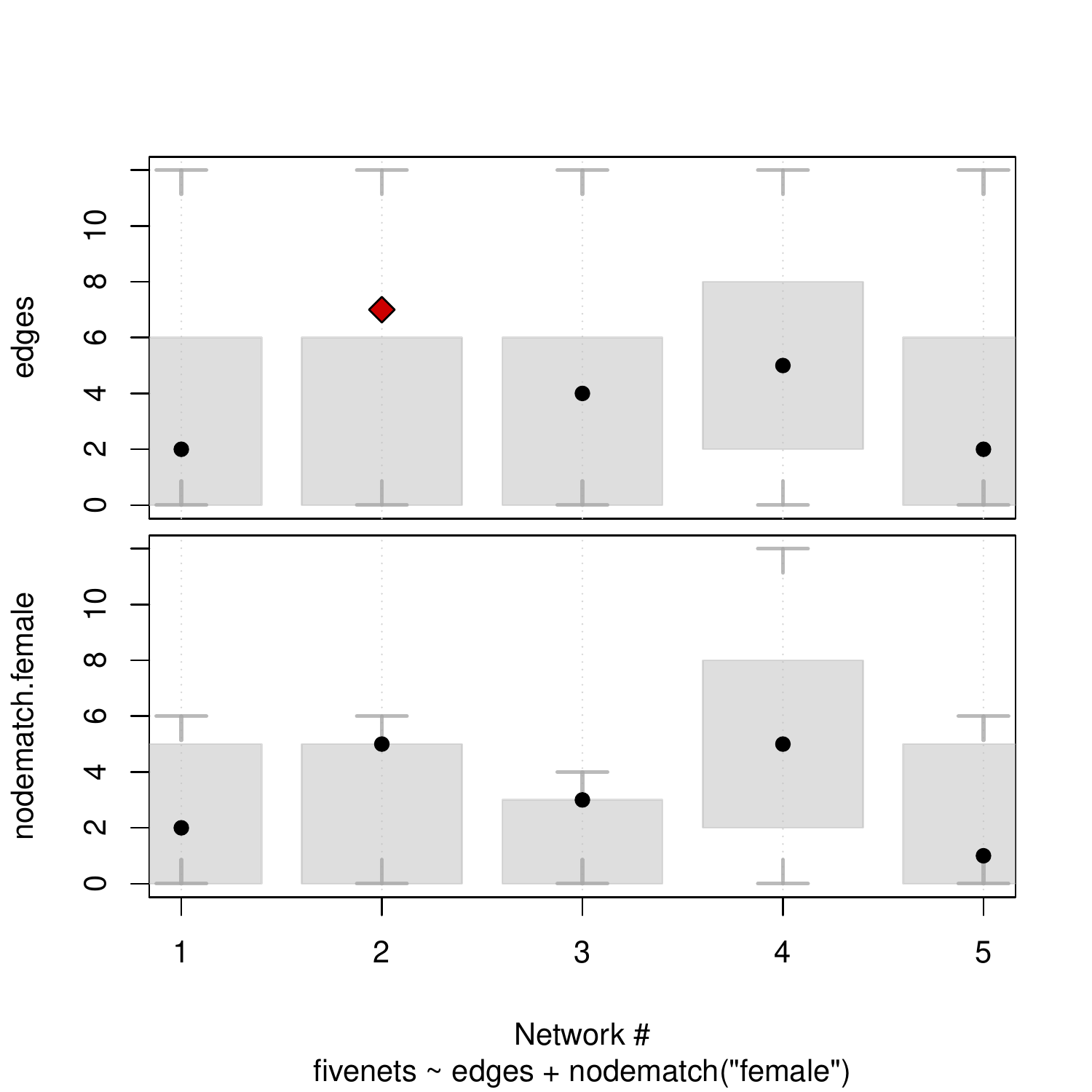}
    \label{fig:fivenets-gof}
\end{figure}

Researchers that apply ERGMs should be familiar with the graphical goodness-of-fit (GOF) diagnostics that are used to assess how well the estimated model can reproduce graphs that are similar to the observed graph on a range of local and global graph statistics \cite{Hunteretal2008}. In the case of \ergmitos{} applied to small networks, local graph statistics will be more relevant than global statistics to assess GOF. For example, the graph geodesic distribution (i.e., the distribution of shortest-path lengths) is often used to assess GOF for larger networks, but this is clearly less relevant in the case of small networks (like in our case, containing at most 6 nodes if directed, or 8 if undirected) because the shortest-path length between any two nodes typically lies between one and three steps. Therefore, we focus the GOF analysis on the parameters fit in the model as the minimum set of local graph statistics, as shown in \autoref{fig:fivenets-gof}; and depending on the model complexity a more comprehensive set of local statistics may be needed. An important difference in our approach compared to traditional GOF assessments for ERGMs is that we are able to enumerate the full support of the model, and so instead of showing a boxplot we present a 90\% exact confidence interval per-statistic per network, comparing the fitted model's distribution with the observed parameters. % A detailed discussion of this aspect of the \ergmitos{} is presented at the end of this paper (\sectionSection XX).

\begin{figure}[tb]
    \centering
    \caption{Surface of the log-likelihood function of the pooled \ergmito{} model. Lighter colors represent higher values while darker ones represent lower values. The red dot corresponds to the location of the MLE estimate of the model.}
    \includegraphics[width=.7\linewidth]{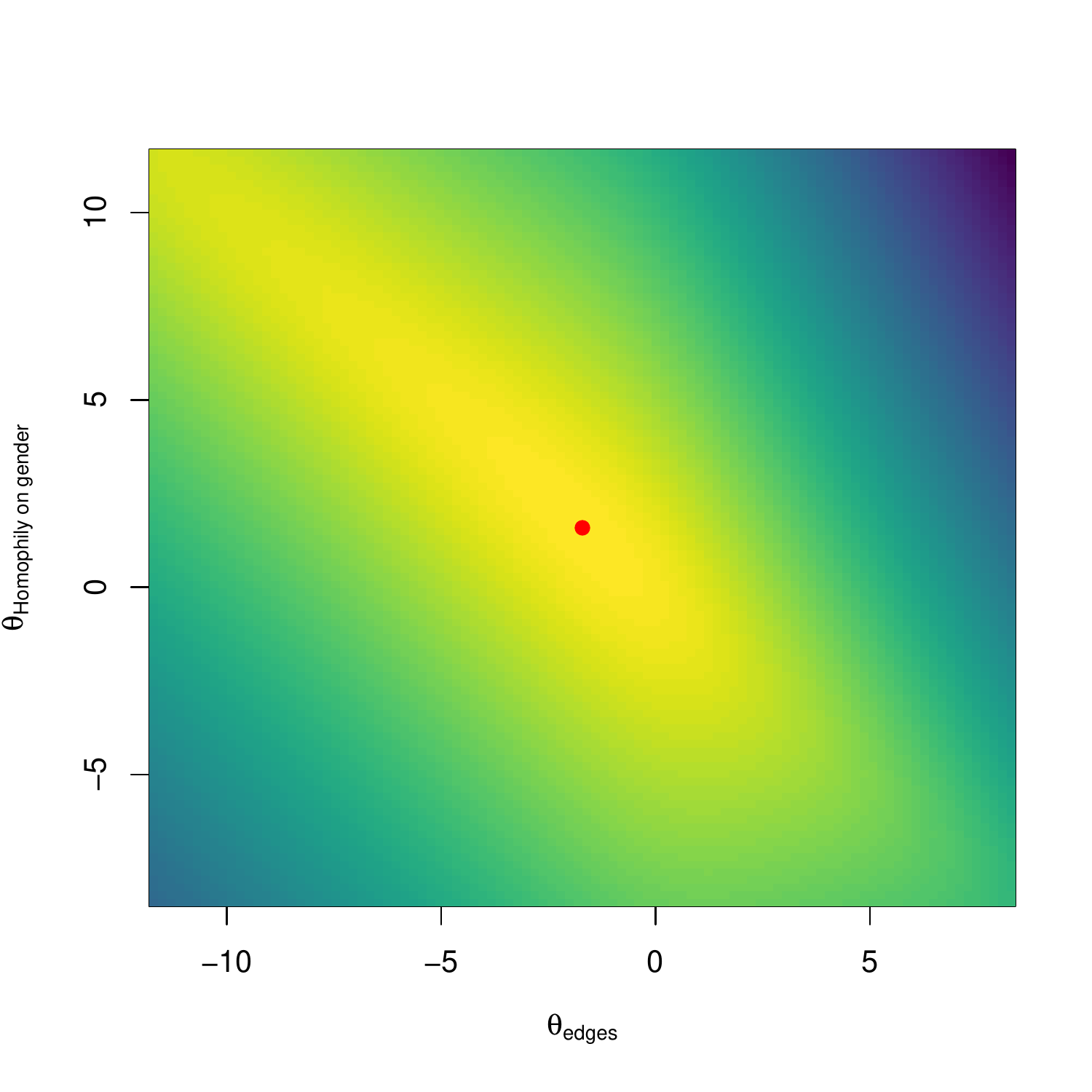}
    \label{fig:fivenets-loglike}
\end{figure}

An important advantage of the \ergmitos{} over ``regular'' ERGMs is that we can observe the surface of the log-likelihood over different combinations of parameters in a rather straightforward way. This, together with the GOF analysis should be a routine step done after every \ergmito{} fit. \autoref{fig:fivenets-loglike} shows the surface of the log-likelihood function around the solution parameters to the maximization problem.

The ability to calculate the surface of the exact likelihood function provides additional tools for assessing the quality of the estimated set of parameters. One good use of this diagnostic is to evaluate the roughness of the log-likelihood function, which in principle should give us an idea of the likelihood of the maximization process failing to reach a global maxima, or estimates being close to problematic (e.g., generating empty or fully connected graphs) areas of the parameter space.

\section{Simulation study\label{sec:simulation-study}}

We conducted two sets of simulations where we compare the performance of the Maximum Likelihood Estimator [MLE] with that of the Monte Carlo MLE [MC-MLE] and Robbins-Monro Stochastic Approximation [RM] in terms of bias, power, type I error rates, and overall computation time. In the first set of simulations, we analyze empirical bias, empirical power, and overall computation time of each estimator in a scenario where the ERGM is defined by \textit{edgecounts} and \textit{transitive triads}. For the second set of simulations, we look at empirical type I error rates when ERGMs are mis-specified by including a \textit{transitive triad} term in the context of a data-generating-process that only includes an \textit{edgecount} statistic.

The code used to reproduce this entire section can be found at %BLIND REVIEW.
\url{https://github.com/muriteams/ergmito-simulations}.

\subsection{\label{subsec:design}Empirical Bias and Power}

Using the \ergmito{} R package, we generated 20,000 samples (datasets), with each sample consisting of several small networks defined by the parameters $edges$ (edgecount) and $ttriads$ (number of transitive triads). Each sample was generated using different combinations of parameters. While all come from an ERGM model defined by edgecounts and number of transitive triads, for every sample we specified: (1) population parameters for the ERGM, (2) the size of the sample (i.e., the number of networks in the sample), and (3) the composition of the sample in terms of the combination of networks of size four and five. A detailed description of each one of these three components used to draw the samples follows:

\begin{enumerate}
\item \textbf{Population parameters}: First we drew two numbers from a piece-wise Uniform distribution with values in $[-2, -.1]\cup[.1, 2]$, ($\params_{edges}, \params_{ttriads}$), which corresponded to the parameters associated to the statistics \textbf{edgecount} and \textbf{number of transitive triads}. This specifies the ERGM from which we will draw the networks from. This is akin to the approach taken by \cite{Schweinberger2015} , although we took a more conservative approach than their ranges of (-5,0) and (0, 5) for the parameters ``edges'' and ``triangles'' in order to increase the number of \textit{irrelevant} draws (i.e., samples composed mostly of either empty of fully connected graphs, or networks with no transitive triads).

\item \textbf{Number of networks per sample} Then, we specified the number of networks to generate from the models defined in the previous step, using one of the following sample sizes $\{5, 10, 30, 50, 100, 150, 200, 300\}$. The 20,000 simulations were equally split across the various sample sizes (i.e., the simulation study was based on 2,500 samples comprised of 5 networks; 2,500 samples comprised of 10 networks, etc.)

\item \textbf{Number of nodes per network} Finally, the composition of each sample, in terms of the number of nodes that each network has, was uniformly-random selected from the pairs $\{N, 0\}, \{N - 1, 1\}, \dots, \{1, N - 1\}, \{0, N\}$, where the first number of each pair is the number of networks of size 4, and the second is the number of networks of size 5 in the sample. As an example, if the sample size selected in the previous step was 30, then the possible pairs to select from would be $\{30, 0\}, \{29, 1\}, \dots, \{1, 29\}, \{0, 30\}$, so that samples in which all networks were of size 4 (meaning we draw the pair $\{30, 0\})$ or size 5 (again, selecting the pair  $\{0, 30\}$) were equally likely. 
\end{enumerate}

For each one of the 20,000 simulated datasets, we then estimated the model using MLE, as implemented in the \textit{ergmito} R package, %\cite{vegayon2018}
and MC-MLE and RM, as implemented in statnet's \textit{ergm} R package \cite{Handcock2018,hunter2008}. In the case of the latter two, the pooled estimation was done by fitting what is known in the literature as a block-diagonal model in which (a) networks are stacked together in a single adjacency matrix, and (b) the sampling space for the MCMC process is constrained to sample from graphs where ties are only possible within blocks. In the case of the MCMC estimator, we set the control parameters \textit{interval} and \textit{samplesize} to 2,048, with a burn-in of 2,048 x 16 = 32,768; all double the of the current default values specified in the \textit{ergm} package, so that we could increase the precision of our estimates. And in  cases where the algorithm failed to return any estimates, we increased the control parameters \textit{interval} and \textit{samplesize} to 10,000.

\subsubsection{Analysis preface}

After simulating the data and estimating the models, we found that there were several cases in which the programs implementing the three algorithms did not converge, and either returned estimates with a warning to the user, or failed without returning a meaningful message to the user. First, the MLE implementation in \textit{ergmito} had zero failures, meaning that, even if the optimization failed to converge, the program provided the user with a meaningful report in all cases. Second, while the MC-MLE implementation of the \textit{ergm} package did fail without returning \textit{any} form of results in some cases (97 of the 20,000), in each of these instances the program provided the user with a meaningful report of what caused the error. Third, in the case of the Robbins-Monro algorithm [RM], as implemented in the \textit{ergm} package, we observed a high error rate: in about 25\% of the samples, the \textit{ergm} function failed during the estimation process, and returned an uninformative error message to the user (``\textit{NA/NaN/Inf in foreign function call (arg 13)}''). This error rate should be interpreted with some context; the implementation of the RM algorithm has received less attention, and thus less optimization, that the MC-MLE method. While the \textit{PNet} \cite{wang2006pnet} software provides a more mature implementation of the RM algorithm, we chose to use \textit{statnet}'s implementation as it was better suited for the implementation of our simulation study.
Table \ref{tab:error-sampsize} shows the number of errors as a function of sample size (number of networks) for each estimation method.

% latex table generated in R 3.6.3 by xtable 1.8-4 package
% Sun Apr 26 23:39:32 2020
\begin{table}[htb]
\centering
\begin{tabular}{lccc}
\toprule & \multicolumn{3}{c}{\# of errors} \\ \cmidrule(r){2-4}
Sample size & MLE & MC-MLE & RM \\ 
  \midrule
5 &   0 &  44 & 1,274 \\ 
  10 &   0 &  21 & 1,058 \\ 
  30 &   0 &  10 & 760 \\ 
  50 &   0 &   3 & 668 \\ 
  100 &   0 &   6 & 583 \\ 
  150 &   0 &   3 & 507 \\ 
  200 &   0 &   4 & 508 \\ 
  300 &   0 &   6 & 460 \\ 
  \midrule Total &   0 &  97 & 5,818 \\ 
   \bottomrule
\end{tabular}
\caption{\label{tab:error-sampsize}Number of times the program failed to fit a model and returned with an error. This shows the overall error rate over the full set of 20,000 simulated samples. All but 3 errors of the RM implementation happened on cases where the sufficient statistics were on the boundary.} 
\end{table}

 Nearly all of the errors (cases in which the software failed and returned with an error) observed in RM, all but three occur on realizations of the data-generating-process that yielded uninteresting cases, where either of the observed sufficient statistics was on the boundary of their support, e.g. fully connected graphs or graphs with no triads.

\begin{figure}[htb]
 	\centering
 	\caption{\label{fig:failed}Distribution of the average sufficient statistics per sample. Since samples can contain networks of sizes four and five, we have re-scaled the sufficient statistics counts by each network size's corresponding maximum value so these range from zero to one. Most of the cases in which methods failed to converge happened in scenarios where either all the graphs in the sample were fully connected or there was no transitive triad; exactly the cases that we excluded for the reminder of the analysis.}
 	\includegraphics[width=.8\linewidth]{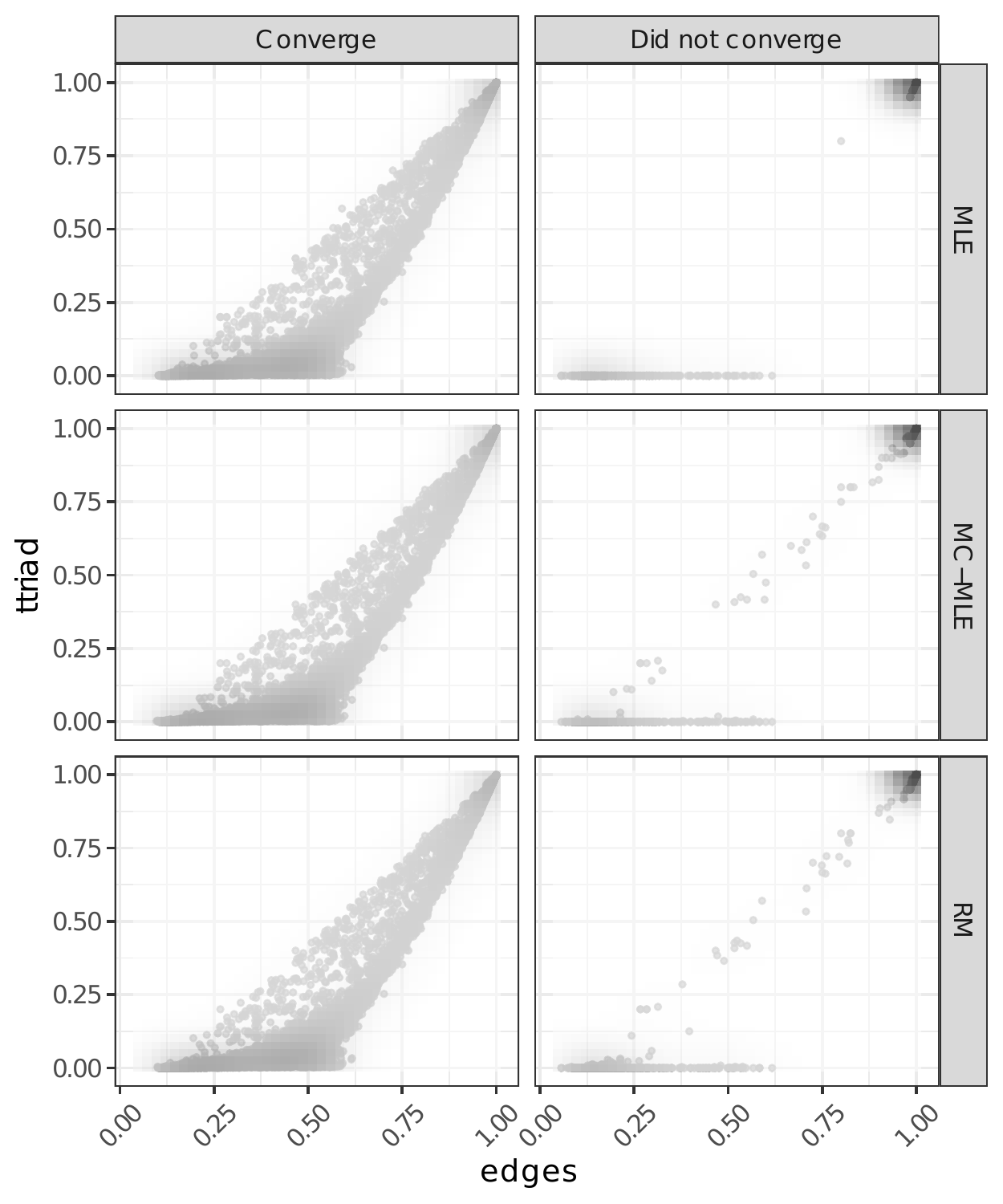}
\end{figure}

With respect to those cases in which the algorithm failed to converge (which includes both software errors and the program reporting lack of convergence), \autoref{fig:failed} shows the distribution of the sufficient statistic split based on whether the algorithm converged or failed to do so. As shown in the figure, when the algorithms did not converge it was typically due to sufficient statistics falling on the boundary of its support (convex-hull problem). This was especially true for the case of the MLE implementation of \textit{ergmito}, as all but one of the convergence failures were on the boundary. While MLEs can be obtained in some of those cases (see appendix \ref{sec:evaluation-of-estimates} and \cite{Handcock2003}), in general, estimating such models has no practical utility. We therefore focused our analysis on samples of networks for which the aforementioned model is appropriate: all subsequent analyses include only those data sets where the observed sufficient statistics, \textit{edgecounts} and \textit{number of transitive triads}, were not on the boundary of its support for \textit{at least} one network in the sample. In other words, we \textit{included} the sample if it: (a) had at least one graph that was not fully connected, and (b) had at least one transitive triad in at least one network. Of the 20,000 simulated data sets, 14,185 met the criteria. 

Overall, practitioners should bear in mind that the cause of errors that arise during the estimation process can be based on the method or software, and when this is captured by the program it can be informative to both users and developers.  

\subsubsection{Empirical Bias and Power}

\begin{figure}[htb]
	\centering
	\caption{\label{fig:bias}Empirical distribution of the bias per model parameter, for MC-MLE and MLE estimation methods. In general we see that the parameter estimates' bias is centered around zero and both MC-MLE (ERGM) and MLE (ERGMito) have about the same bias in our simulation study.}
	\includegraphics[width=.9\linewidth]{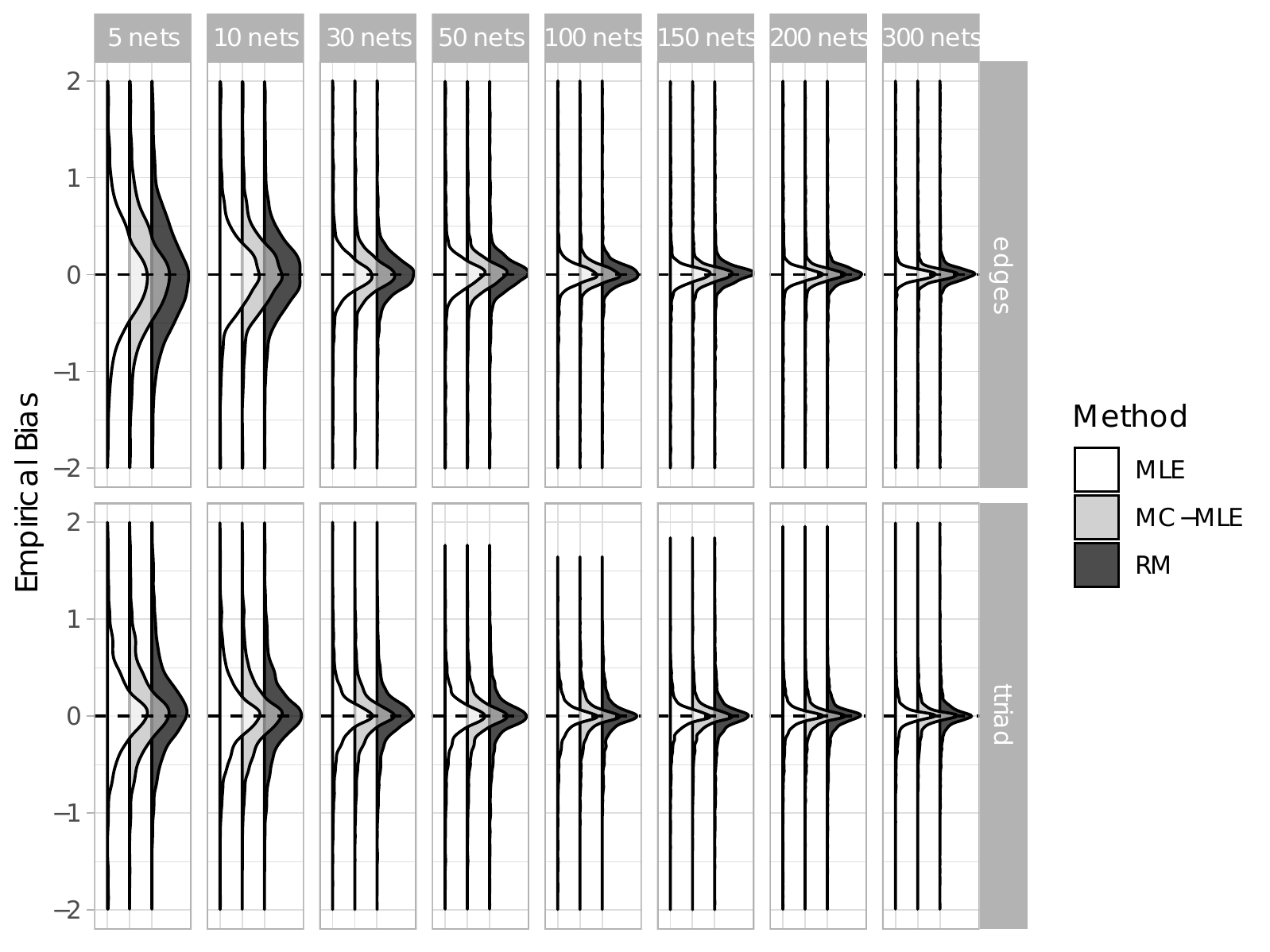}
\end{figure}

As shown in \autoref{fig:bias}, all three estimation methods behaved similarly in terms of empirical bias in the models studied here. As the size of the sample of networks in the dataset increased (i.e., when there were more networks within the sample), the empirical bias of all three, MLE, MC-MLE and RM, decreased, as expected.

Looking closer at the biases, we noticed that, while all methods show some kind of bias, MLE has (on average) the smallest. As showed in \autoref{tab:empirical-bias-sim}, at the 95\% confidence level, all three methods tend to overestimate the \textit{edges} parameter. On the other hand, with the exception of the RM method, both MLE and MC-MLE tend to underestimate the \textit{transitive triads} parameter; yet, the RM method has the widest confidence interval for that parameter.

\begin{table}[ht]
\centering
\begin{tabular}{rccc}
  \toprule
 & MLE & MC-MLE & RM \\ 
  \midrule
edges & [0.27, 0.36] & [1.23, 1.65] & [0.55, 1.54] \\ 
  ttriads & [-0.05, -0.03] & [-0.22, -0.16] & [-0.15, 0.48] \\ 
   \bottomrule
\end{tabular}
\caption{\label{tab:empirical-bias-sim}Empirical bias. Each cell shows the 95\% confidence interval of each methods' empirical bias.} 
\end{table}
    
\begin{figure}[tb]
	\centering
	\caption{\label{fig:power}Empirical power by dataset size and effect size (the later considering only magnitude), for ERGM and \ergmito{} estimation methods. Power increases for both MC-MLE (ERGM) and MLE (\ergmito{}) with increases in the size of the dataset and effect size. There are indistinguishable differences in power between the two estimation methods.}
	\includegraphics[width=.9\linewidth]{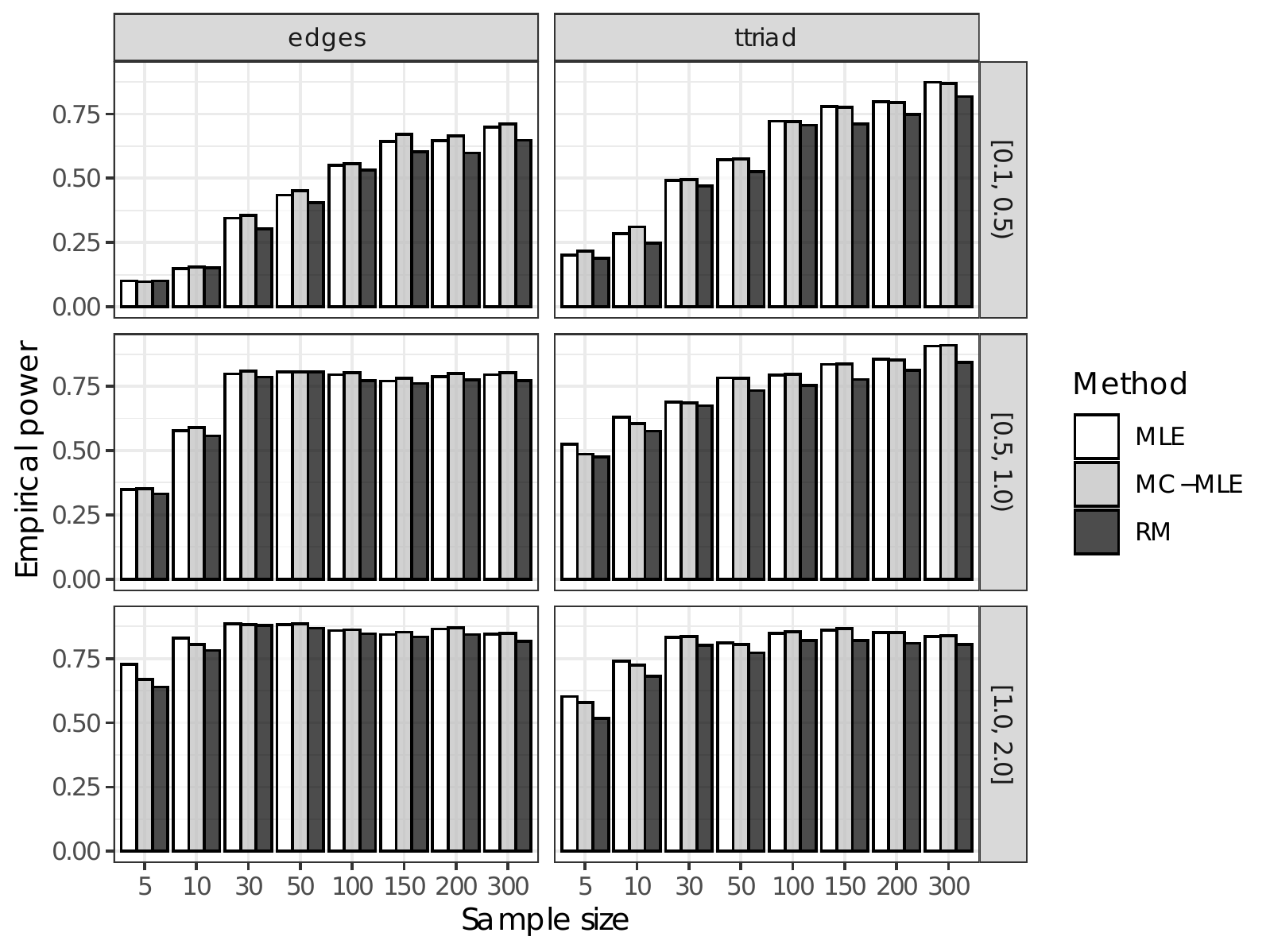}
\end{figure}

Empirical power levels, calculated as the proportion of times that the method reported a significant effect at the 5\% level in the same direction as the data-generating-process parameter, is depicted in \autoref{fig:power}. For each method, a single bar in the figure shows the empirical power level for the corresponding combination of sample size (x-axis), parameter (columns), and effect size (rows). There are three main findings to highlight: first, as expected, power increases as both sample size and effect size increase; second, both MLE and MC-MLE behave very similarly with no statistically significant differences across sample and effect size; and third, compared to MLE, RM had a statistically significant smaller power level at various sample and effect sizes combinations, with the largest differences observed on transitive triads Although there may be some inherent properties of each method that may benefit MLEs, this again may be due to less emphasis on the implementation of RM in the  \textit{ergm} package. Finally, as an anecdotal observation, it is interesting to see that, in the case of effect sizes of magnitude [0.5, 1.0), the discovery rate for the \textit{ttriads} parameter reaches nearly 0.75 for sample sizes between 30 to 50 networks, which is a rather common sample size in the study of small networks such as teams, families, and sometimes ego-networks. 

\begin{figure}[tb]
	\centering
	\caption{\label{fig:power-prop5s}Empirical power by proportion of networks of size five in the sample (color coded) and sample size (rows).}
	\includegraphics[width=.8\linewidth]{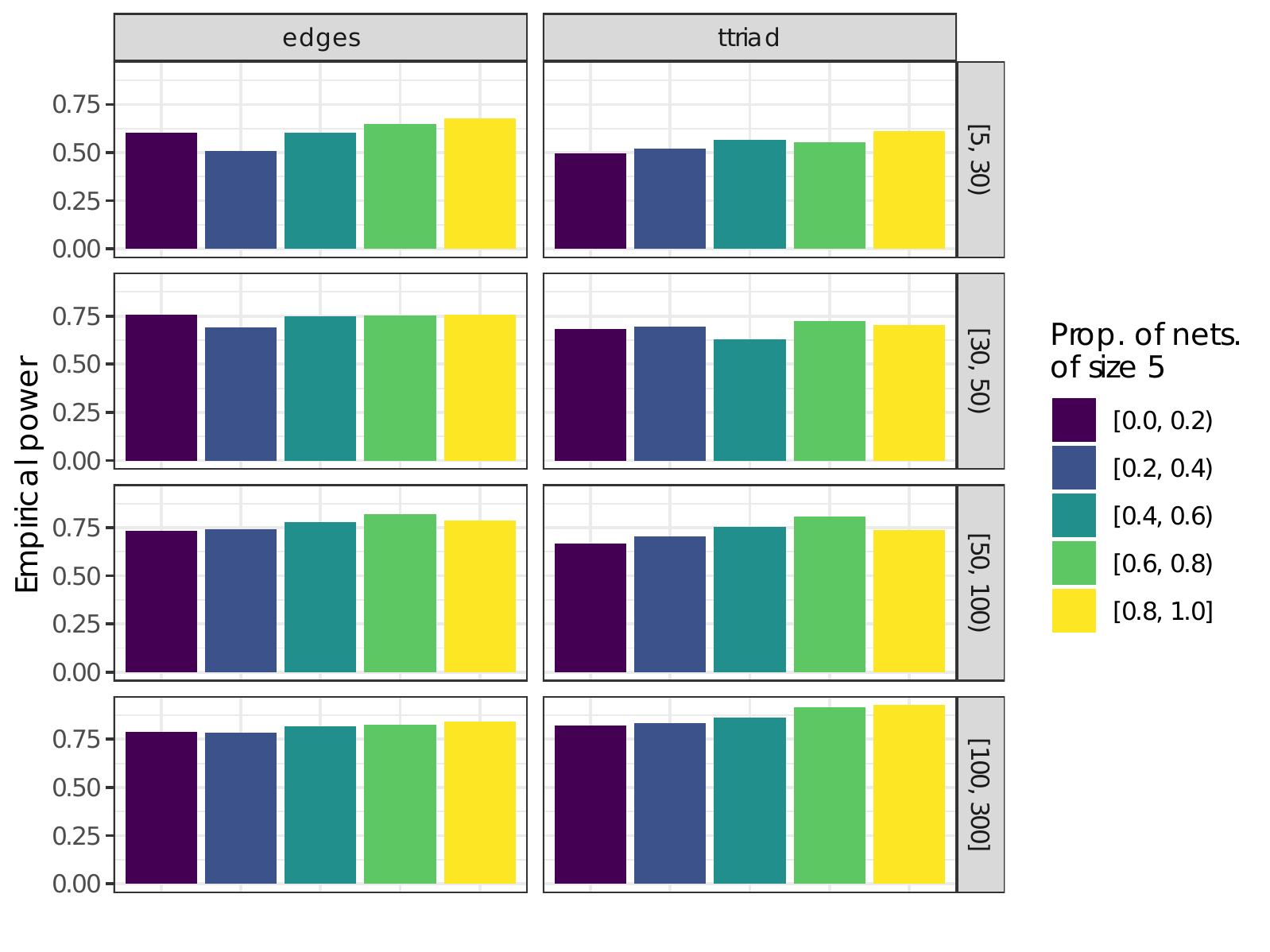}
\end{figure}

\autoref{fig:power-prop5s} shows the effect of the composition of the sample in each dataset, in terms of the proportion of networks of size five (vs. size four), and the number of networks per dataset. In this case, we observe no meaningful patterns that would indicate the dataset composition is related to power.

\begin{figure}[tb]
	\centering
	\caption{\label{fig:elapsedtime}Distribution of elapsed time (in seconds) for the estimation process for MC-MLE (ERGM) versus MLE (using \ergmito{}). Overall, the MLE implementation is orders of magnitude faster compared to the time required by the MC-MLE implementation to do the parameter estimation.}
	\includegraphics[width=.9\linewidth]{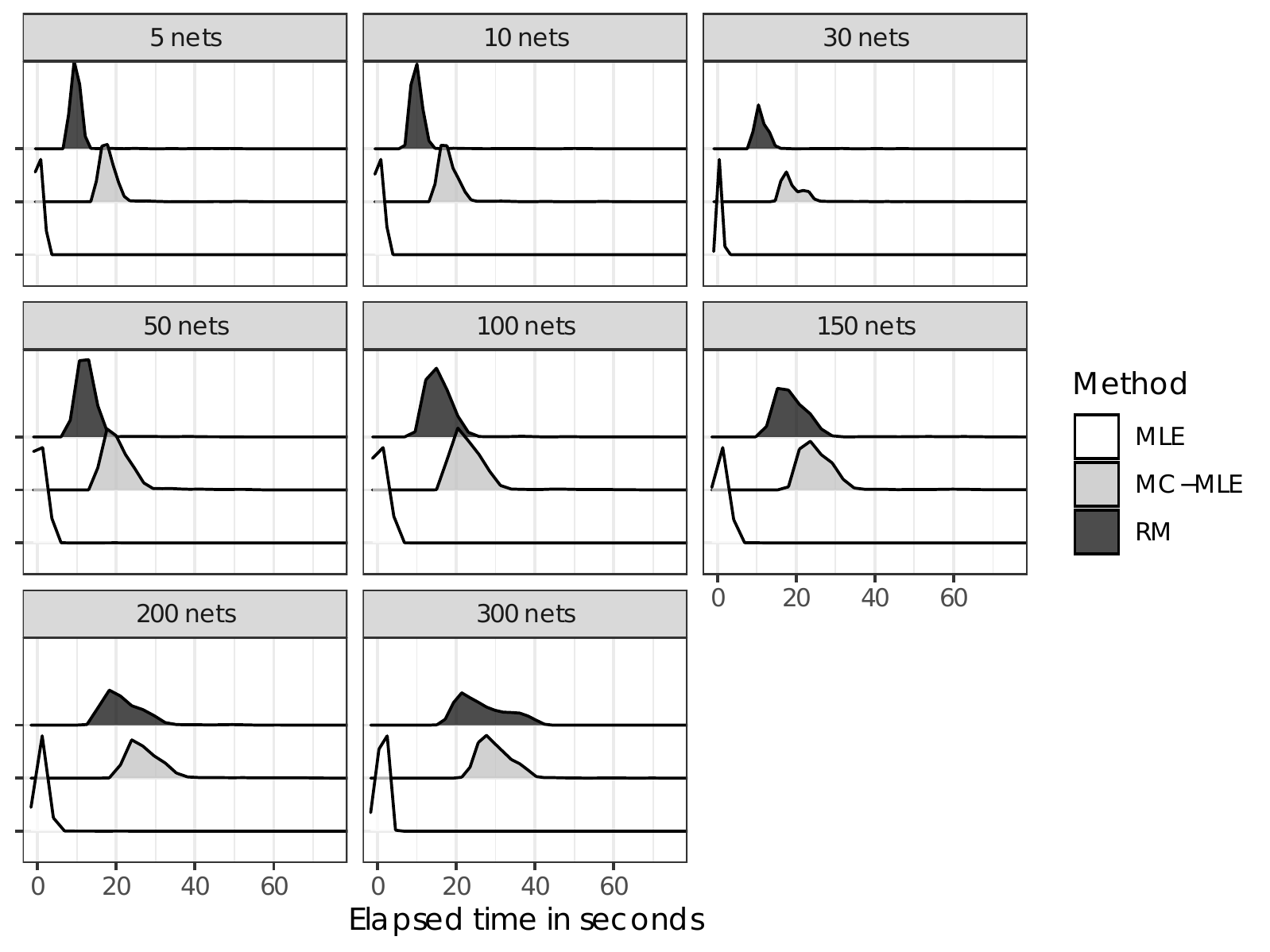}
\end{figure}

One remarkable difference between the three estimation methods featured by the simulations is the overall computing time needed to fit the models. While the computation of exact likelihoods and gradients is still very computationally intensive, the total time needed to obtain MLEs is still  significantly less than what is needed to by the other two methods. As shown in \autoref{fig:elapsedtime}, MLE can be orders of magnitude faster than MC-MLE and RM. Therefore, while all three estimators show very similar properties in terms of power and bias, practitioners will benefit by using MLE when modeling small networks because it may substantially reduce computation time.

Nevertheless, while MLE is generally faster than the other two methods, there are some scenarios in which the speed gains may not be as dramatic as those shown here. The biggest computational bottleneck that the MLE estimation faces is the calculation of the full support of the sufficient statistics. In the case of structure-only statistics, \textit{ergmito}, and actually \textit{ergm}, computes the full distribution very quickly, but, as the model starts to become more complex, such calculation becomes more and more expensive. Yet, once the full enumeration of the support of sufficient statistics is done, finding MLEs becomes \textit{trivial}, making the implementation of other statistical tools such as bootstrap or forward/backward model selection feasible to implement. Bootstrapping of ERGMs is illustrated in \autoref{sec:empirical}.

\subsection{Type I error rates}

Using the same procedure described in \autoref{subsec:design}, we simulated 35,000 datasets comprised of Bernoulli networks (i.e., an ERGM model only defined by the \textit{edgecounts} sufficient statistic). In this case, we drew  different sets of sample sizes: for each of $\{5, 10, 15, 20, 30, 50, 100\}$ we generated 5,000 datasets using the Bernoulli model with \textit{edgecount} parameter uniformly distributed in the range $[-2, -.1]\cup [.1, 2]$. We then estimated the models using MLE, MC-MLE, and RM and calculated the type I error rates using a misspecified model; that is, fitting ERGMs that included a \textit{transitive triads} count statistic. As with the previous simulations, we only analyze datasets that either had at least one not fully connected graph and had at least one transitivite triad in at least one network. Fortunately, as \autoref{tab:typeI} shows, most of the cases did.

Table \ref{tab:typeI} shows the type I error rates per sample size for each of the three methods. In general, MLE report lower error rates compared to MC-MLE and RM, when models were fit to datasets with sample sizes of 20 or fewer networks, the MLE had a better performance than MC-MLE as it reported smaller type I error rates that were much closer to the nominal 5\% level. Datasets with 30 or more networks had no significantly different type I error rates between the two methods. Compared to RM, the simulation study shows MLE has a better performance when estimating pooled-data models with 10 or less networks. No significant difference is observed when dealing with samples of 15 or more networks.

% latex table generated in R 3.6.3 by xtable 1.8-4 package
% Wed Apr 29 12:49:06 2020
\begin{table}[ht]
\centering\small
\begin{tabular}{*{2}{m{.1\linewidth}<\centering}*{5}{c}}
\toprule & & \multicolumn{3}{c}{P(Type I error)} & \multicolumn{2}{c}{$\chi^2$ (vs MLE)}\\ \cmidrule(r){3-5} \cmidrule(r){6-7}
Sample size & N. Sims. & MLE & MC-MLE & RM & MC-MLE & RM \\ 
  \midrule
5 & 4,325 & 0.066 & 0.086 & 0.086 & 11.36 *** & 11.36 *** \\ 
  10 & 4,677 & 0.063 & 0.078 & 0.073 & 8.44 *** & 3.73 * \\ 
  15 & 4,818 & 0.060 & 0.072 & 0.063 & 5.50 ** & 0.41  \\ 
  20 & 4,889 & 0.054 & 0.065 & 0.061 & 5.30 ** & 2.05  \\ 
  30 & 4,946 & 0.053 & 0.059 & 0.055 & 1.60  & 0.07  \\ 
  50 & 4,987 & 0.053 & 0.055 & 0.047 & 0.16  & 1.67  \\ 
  100 & 4,999 & 0.054 & 0.054 & 0.050 & 0.00  & 0.81  \\ 
   \bottomrule
\end{tabular}
\caption{\label{tab:typeI}Empirical Type I error rates. The $\chi^2$ statistic is from a 2-sample test for equality of proportions, and the significance levels are given by *** $p < 0.01$, ** $p < 0.05$, and * $p < 0.10$.} 
\end{table}

\section{Extended Application: The role of gender-homophily on the formation of small teams\label{sec:empirical}}

In this final section, we apply the \ergmitos{} framework to a set of observed social networks in an experimental setting. The data was generated as part of a study that examined the emergence of social networks in small teams. 

The analytic sample consists of 31 small mixed-gender teams that include either four (17 teams) or five members (14 teams). Individuals recruited for the study were University students, participating for research credit or compensation, who were assigned to the teams with two conditioning factors: (1) they did not know the other teammates, and (2) there must be at least one team member who identified as male, and one who identified as female. On average, 55\% of each team's members were female, with no statistically significant difference between the teams (test of equal proportions) nor within the teams when compared to a null of 0.5 (exact binomial test). Each team met face-to-face in a laboratory setting to complete about one hour of group tasks. Immediately after the completion of the group tasks, the team networks were measured using name generators administered in an online survey (that was completed in the lab). \textit{Advice seeking} was one relationship measured, via the question ``\textit{Who did you go to for advice, information, or help to complete the group tasks?}'', and participants could select as many or as few teammates as they liked. These data were used to generate directed graphs that represent the advice-seeking network in each team, where $\adjmat_{ij} = 1$ if $i$ identified $j$ as someone they sought advice from. %The subsequent analyses apply the \ergmitos{} framework to test hypotheses about the social processes that predict the team advice networks that emerged. 

One research question of interest in the field of team science is what is the role of gender and gender-based homophily (i.e., the preference for individuals to form social ties with teammates who match them on gender) in the formation of team networks. Using the \textit{ergmito} R package to model the team advice networks and test hypotheses about gender and network dynamics, we illustrate how exact calculation of ERGM likelihoods can be leverage to go beyond traditional ERGM analysis. Overall, the analysis consists of two parts: (1) building a baseline model that only includes structural features of the graph, and (2) using that model to test if gender-homophily is a prevalent feature of the data, while also controlling for other gender-based terms in a multivariate fashion. 

In the structural-terms-only model, we fitted five different models based on the following terms:

\begin{itemize}
    \item \textbf{Edge count} (edges): This accounts for the overall density of the graph and is usually compared to that of a constant term in regression analyses. This is calculated as $\sum_{ij}\adjmat_{ij}$.
    \item \textbf{Number of transitive triads} (ttriads): This statistic, also known as balanced triangles or transitive triples, captures the phenomenon of social clustering and balance; where ``\textit{the friend of my friend is my friend}''. In this context it indicates that ``\textit{the advisor of my advisor is my advisor}''. This term is calculated as follows: $\sum_{i}\sum_{j<k}\adjmat_{ij}\adjmat_{jk}\adjmat_{ik}$.
\end{itemize}

To illustrate the flexibility of estimating ERGMs with the \textit{ergmito} R package, we generated three additional terms to be included in the models using the \textit{edges} and \textit{ttriads} terms. First, we included two interaction effects, one per term, with an indicator variable which equals to one if the corresponding network was of size five, and zero if it was size four. We also added an offset term as that proposed by \cite{Krivitsky2011} which has the nice property of being size-invariant; i.e., it preserves the mean degree as the network size increases. All of these additional terms allowed us to control for differences as a function of the network size. A valuable benefit of these additional statistics is that users can add interaction effects or variable transformations to the models; a feature that, currently, is not easily achieved in other available frameworks (see for example \cite{Hunter2013,Handcock2019}). Just like we showed earlier in \autoref{tab:obs-suff-fivenets}, \autoref{tab:example-suffstats} shows an example of the target statistics used in the models for 6 of the 31 networks (i.e., the array of observed sufficient statistics). With these five statistics we estimated five different models, including a bootstrapped version of the one with the best overall fit. \autoref{tab:ci-ergm-baseline} shows the results.

% latex table generated in R 3.6.3 by xtable 1.8-4 package
% Mon Apr 27 23:32:51 2020
\begin{table}[ht]
\centering
\begin{tabular}{*{3}{c}*{3}{m{.15\linewidth}<\centering}}
  \toprule
  (1) & (2) & (3) & (4) & (5) & (6) \\
Size ($n$) & edges & ttriads & $\text{edges}\times\isone{n = 5}$ & $\text{ttriads}\times\isone{n = 5}$ & $\text{edges}\times\log{1/n}$ \\ 
  \midrule
  4 &  10 &  14 &   0 &   0 & -13.86 \\ 
    4 &   6 &   2 &   0 &   0 & -8.32 \\ 
    4 &   4 &   0 &   0 &   0 & -5.55 \\ 
    5 &   6 &   1 &   6 &   1 & -9.66 \\ 
    5 &   8 &   8 &   8 &   8 & -12.88 \\ 
    5 &   6 &   2 &   6 &   2 & -9.66 \\ 
    \multicolumn{6}{c}{\dots\textit{25 more rows}\dots} \\
   \bottomrule
\end{tabular}
\caption{\label{tab:example-suffstats}Example of observed sufficient statistics for the team advice networks. Pooled-data ERGMs have multiple observed sufficient statistics (also known as target statistics). Furthermore, as shown here, we can manipulate common statistics as \textit{edges} (2) and \textit{ttriads} (3) to include, e.g. interaction effects (4) and (5), or more complex transformations, e.g. (6).}
\end{table}

\begin{table}[tb]
\footnotesize
\begin{center}
\begin{tabular}{l c c c c c c }
\toprule
 & (1) & (2) & (3) & (4) & (5) & (3b) \\
\midrule
edges                        & $-0.72^{***}$ & $0.73^{***}$ & $-0.53^{***}$ & $-0.85^{***}$ & $-0.56^{*}$  & $-0.53^{***}$ \\
                             & $(0.13)$      & $(0.13)$     & $(0.15)$      & $(0.14)$      & $(0.23)$     & $(0.12)$      \\
ttriad                      & $0.29^{***}$  & $0.33^{***}$ & $0.36^{***}$  & $0.50^{***}$  & $0.38^{***}$ & $0.36^{***}$  \\
                             & $(0.05)$      & $(0.05)$     & $(0.06)$      & $(0.07)$      & $(0.11)$     & $(0.05)$      \\
$\text{edges}\times\isone{n= 5}$          &               &              & $-0.53^{***}$ &               & $-0.49$      & $-0.53^{***}$ \\
                             &               &              & $(0.12)$      &               & $(0.28)$     & $(0.12)$      \\
$\text{ttriad}\times\isone{n= 5}$        &               &              &               & $-0.22^{***}$ & $-0.02$      &               \\
                             &               &              &               & $(0.05)$      & $(0.12)$     &               \\
\textit{offset}\\
\hspace{5mm}$\text{edges} \times \log(1/n)$ &               & Yes       &               &               &              &               \\
                             &               & $$           &               &               &              &               \\
\midrule
AIC                          & 651.38        & 641.02       & 637.28        & 640.40        & 639.26       & 637.28        \\
BIC                          & 659.74        & 649.39       & 649.83        & 652.95        & 655.99       & 649.83        \\
Log Likelihood               & -323.69       & -318.51      & -315.64       & -317.20       & -315.63      & -315.64       \\
Num. networks                & 31            & 31           & 31            & 31            & 31           & 31            \\
Time (seconds)               & 0.55          & 0.99         & 0.74          & 0.76          & 0.74         & 10.12         \\
N replicates                 &               &              &               &               &              & 1000          \\
N Used replicates            &               &              &               &               &              & 1000          \\
\bottomrule
\multicolumn{7}{l}{\scriptsize{$^{***}p<0.001$, $^{**}p<0.01$, $^*p<0.05$}}
\end{tabular}
\normalsize 
\caption{Structural models. Model (2) includes Krivitsky et al (2011) offset term. Besides of the common GOF statistics, the table includes the number of networks used, elapsed time to fit the model, and, in the case of Model (3b) which is a bootstrapped version of model (3), number of replicates fitted and included in the bootstrap variance estimate.}
\label{tab:ci-ergm-baseline}
\end{center}
\end{table}

The results, \autoref{tab:ci-ergm-baseline}, indicate that transitive triads (\textit{ttriads}) were more prevalent than expected by chance; which is common in positive affiliation and collaboration networks. Parameter estimates for the \textit{ttriads} term were also robust with significant and positive effects across the different model specifications. Second, we found that controlling for size of the network mattered. The results of models (3) and (4) show that allowing networks of size 5 to have different parameters associated with number of edges or transitive triads (with networks of size 4 as a reference), significantly improved model fit relative to model (1). Yet, as shown in model (5), these interaction effects were not jointly significant. Regarding model (2), which includes the offset $\text{edges} \times \log{1/n}$, we see that the \textit{edges} parameter flips from negative 0.72, to positive 0.73, which should be interpreted in the context of this offset change. For example, in the case of the Bernoulli model, the probability of an individual tie for a network of size 4 would be $\text{logit}^{-1}(-\log{4} + 0.73) \approx \text{logit}^{-1}(-0.66) \approx 0.34$, i.e. less than 0.5 which is the expected value under the null.

Of the five models, model (3) had the best overall fit, the lowest AIC and BIC, and so it was retained as the structural baseline model for the subsequent analyses. To finalize this first stage of analysis, we calculated the standard errors of model (3) using bootstrap \cite[see][]{Wooldridge2010}; with the results reported in column (3b). This final model had no meaningful changes in standard errors compared to (3); although they were slightly smaller compared to MLEs in (3). Additionally, the elapsed time for this bootstrapping process was negligible: remarkably, we fit 1,000 ERGMs in about 10 seconds, which further highlights how speed and model specification-flexibility are key features of fitting ERGMs using Maximum Likelihood.

\begin{figure}[ht]
    \centering
    \includegraphics[width=.7\linewidth]{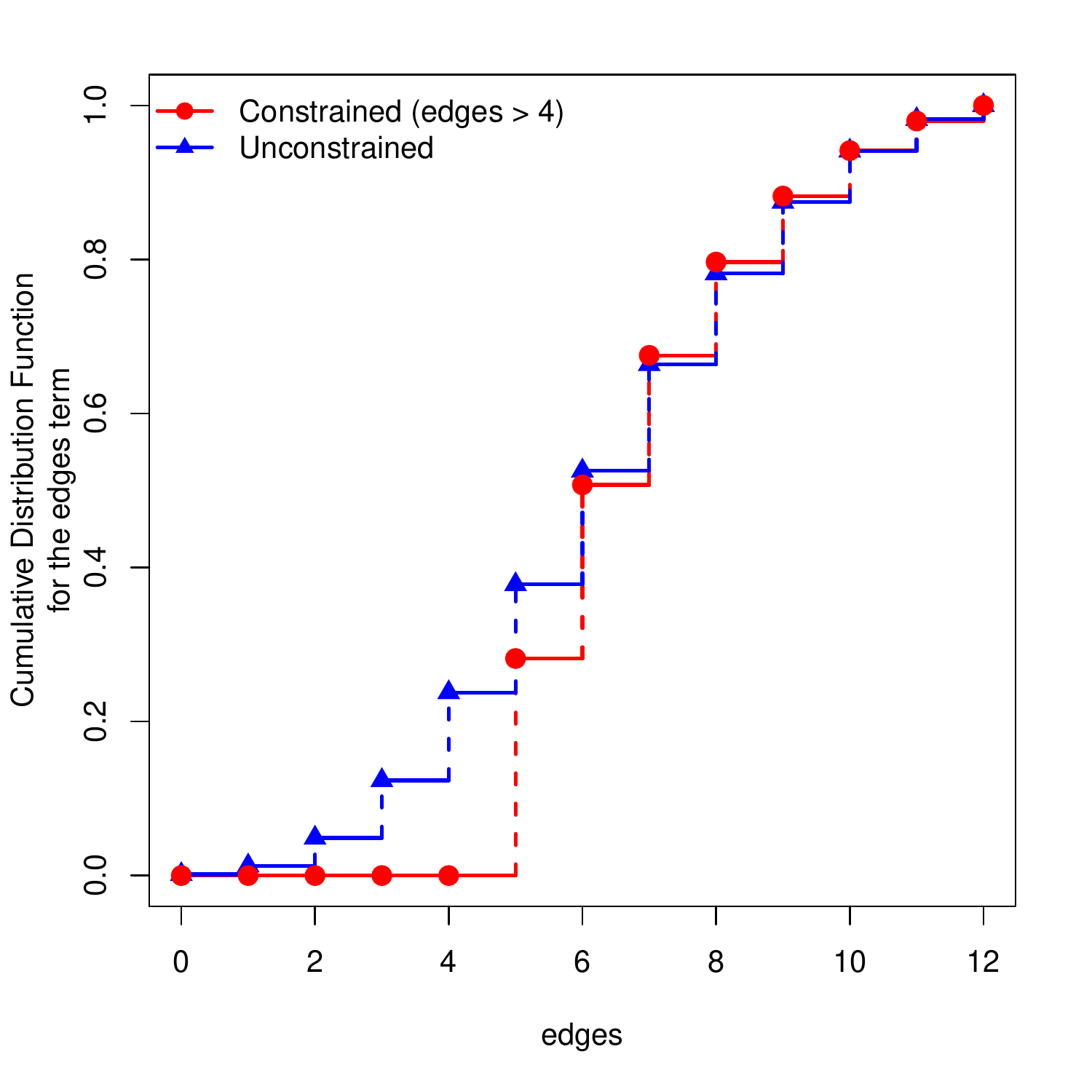}
    \caption{Marginalized Cumulative Distribution Function (CDF) for the \textit{edges} sufficient statistic for a network of size 4 (up to 12 ties). The blue line shows the CDF for the edges term according to model (1) in \autoref{tab:ci-ergm-full}, which was fitted without constraining the support of the sufficient statistics, while the red line shows the CDF of model (2), also in \autoref{tab:ci-ergm-full}, which constrained the support to networks with at least 5 ties. Once again, since we can fully enumerate the support, both CDFs are exact, and thus, not simulated.}
    \label{fig:cdf-constrained}
\end{figure}

The second phase of model specification, which uses model (3) as baseline, focused on evaluating the role of gender and gender-homophily in the advice networks, using the following terms: 

\begin{itemize}
    \item \textbf{Gender homophily}: This term equals to the number of ties in which ego and alter are matched on gender. This was calculated as: $\sum_{ij}\adjmat_{ij}\isone{X_{i} = X_{j}}$, where $X_i$ is one if $i$ is a female, and zero otherwise.
    \item \textbf{Female-sender effect}: This term, also known as attribute-activity effect, captures the propensity of females to send ties. Is is calculated as: $\sum_{ij}\adjmat_{ij}X_{i}$.
    \item \textbf{Female-receiver effect}: This term captures the propensity of females to receive ties. It is calculated as: $\sum_{ij}\adjmat_{ij}X_{j}$.
\end{itemize}
Ttaking advantage of the flexibility that the \textit{ergmito} package, and ultimately, using exact likelihoods provides, we also explored modifying the model by means of transformations and offset terms. First, with the purpose of improving the predictive capability of our model, we included the square root of the count of gender-homophilic ties. Other transformations such as interactions with other terms, or centering around a given constant (for example, some population average) could also be implemented. Second, while not the case in our data, we illustrate a hypothetical scenario where the teams had to have at least 5 ties, and we constrained the support of the sufficient statistics to only include networks with five or more ties. We did this by using an offset parameter that equaled $-\infty$ if the network had four or less ties, and zero otherwise. \autoref{fig:cdf-constrained} illustrates the differences between the Cumulative Distribution Function (CDF) associated with \textit{edges} statistic (the probability of observing up to given number of ties, x-axis) calculated from a model with (red line) and without (blue line) the constrained space.

Using offset terms to constraint the support of the model is not a new thing. The \textit{ergm} package features this capability as well, in addition to specialized algorithms to constrain samplings space. Users can also set offsets to $-\infty$ to forbid some configurations, yet, in the case of \textit{ergmito} combining offset terms with the capability of mixing-transforming variables in the model provides the user with greater flexibility. As we did before, an example of six of the 31 networks is shown in \autoref{tab:example-suff-gender}.

% latex table generated in R 3.6.3 by xtable 1.8-4 package
% Wed Apr 29 15:05:35 2020
\begin{table}[ht]
\centering
\small
\begin{tabular}{c*{4}{m{.18\linewidth}<\centering}}
  \toprule
  (1) & (2) & (3) & (4) & (5) \\
$n$ & Homophily (gender) & Receiver (female) & Sender (female) & $\text{Homophily}^{1/2}$ \\ 
  \midrule
  4 &   3 &   5 &   6 & 1.73 \\
    4 &   1 &   4 &   3 & 1.00 \\
    4 &   3 &   4 &   3 & 1.73 \\
    5 &   2 &   2 &   4 & 1.41 \\
    5 &   4 &   7 &   5 & 2.00 \\
    5 &   3 &   4 &   3 & 1.73 \\
    \multicolumn{5}{c}{\dots\textit{25 more rows}\dots} \\
   \bottomrule
\end{tabular}
\caption{\label{tab:example-suff-gender}Example of observed sufficient statistics for the team advice networks (bis). For the second set of ERGMs, we included gender-based effects: homophily (2), receiver (3), and sender (4). Variable (5) is the square root of variable (2).}
\end{table}

Like in the first round of ERGMs, the standard errors of the  final \textit{best model} were re-calculated using bootstrap. \autoref{tab:ci-ergm-full} shows the results.

\begin{table}[tb]
\centering
\footnotesize
\begin{tabular}{l c c c c c c }
\toprule
 & (1) & (2) & (3) & (4) & (5) & (4b) \\
\midrule
edges                                     & $-0.52^{**}$  & $-0.91^{***}$ & $-0.54^{**}$  & $-0.72^{***}$ & $-0.48^{*}$   & $-0.72^{***}$ \\
                                          & $(0.17)$      & $(0.23)$      & $(0.18)$      & $(0.19)$      & $(0.19)$      & $(0.17)$      \\
ttriads                                   & $0.36^{***}$  & $0.46^{***}$  & $0.37^{***}$  & $0.36^{***}$  & $0.36^{***}$  & $0.36^{***}$  \\
                                          & $(0.06)$      & $(0.06)$      & $(0.06)$      & $(0.06)$      & $(0.06)$      & $(0.05)$      \\
Homophily (gender)                        & $-0.03$       & $-0.01$       & $-0.20$       & $-0.12$       & $-0.01$       & $-0.12$       \\
                                          & $(0.20)$      & $(0.21)$      & $(0.46)$      & $(0.20)$      & $(0.20)$      & $(0.20)$      \\
$\text{edges} \times \isone{n = 5}$       & $-0.53^{***}$ & $-0.47^{**}$  & $-0.52^{***}$ & $-0.53^{***}$ & $-0.53^{***}$ & $-0.53^{***}$ \\
                                          & $(0.12)$      & $(0.16)$      & $(0.13)$      & $(0.13)$      & $(0.12)$      & $(0.13)$      \\
$(\text{Homophily})^{1/2}$                &               &               & $0.54$        &               &               &               \\
                                          &               &               & $(1.32)$      &               &               &               \\
Sender (female)                           &               &               &               & $0.46^{*}$    &               & $0.46^{*}$    \\
                                          &               &               &               & $(0.18)$      &               & $(0.18)$      \\
Receiver (female)                           &               &               &               &               & $-0.08$       &               \\
                                          &               &               &               &               & $(0.18)$      &               \\
\textit{Constraint (offset)} \\
\hspace{5mm}$\text{edge} > 4$        &               & Yes        &               &               &               &               \\
                                          &               & $$            &               &               &               &               \\
\midrule
AIC                                       & 639.26        & 569.93        & 641.08        & 634.68        & 641.07        & 634.68        \\
BIC                                       & 655.99        & 586.66        & 661.99        & 655.59        & 661.98        & 655.59        \\
Log Likelihood                            & -315.63       & -280.96       & -315.54       & -312.34       & -315.53       & -312.34       \\
Num. networks                             & 31            & 28            & 31            & 31            & 31            & 31            \\
Time (seconds)                            & 2.26          & 2.32          & 2.28          & 5.10          & 5.19          & 83.97         \\
N replicates                              &               &               &               &               &               & 1000          \\
N Used replicates                         &               &               &               &               &               & 1000          \\
\bottomrule
\multicolumn{7}{l}{\scriptsize{$^{***}p<0.001$, $^{**}p<0.01$, $^*p<0.05$}}
\end{tabular}
\caption{Testing for gender homopholy. Models (1) through (3) include either an interaction term or a transformation of the term \textit{Homphily (gender)}. Models (4) and (5) include female sender and receiver effects, while model (4b) is a bootstrapped version of model (4). Model (2) constraints the sample space by setting an offset restricting the support to networks with at least 5 edges. Furthermore, since three of the 31 teams had less than five ties, these were excluded from the analysis, hence (2) includes 28 of the 31 available networks.}
\label{tab:ci-ergm-full}
\end{table}

As illustrated in \autoref{tab:ci-ergm-full}, in our first three specifications we found no evidence that gender-homophily was a prevalent feature of the advice networks, as the baseline (1), its constrained version (2), and the baseline including a transformed version of gender-homophily (3) failed to reject the null $\theta_{\text{Homophily}} = 0$. Of the other gender-based effects, only the female-sender effect, model (4), was significant. With a coefficient equal to 0.46, the model indicates that, compared to males, females tended to nominate more of their team members as people they sought advice from. Furthermore, we found that the term \textit{Sender (female)} was a confounder of gender-homophily, with the latter changing from -0.03 in model (1), to -0.12 when the female-sender effect is included. Overall, these final models indicate that the team networks are best explained by preferences for balanced advice-seeking triads, and a tendency for females to seek advice from more of their teammates, compared to males. 

With Model (4), \autoref{tab:ci-ergm-full}, having the best fit overall (smallest AIC and BIC), we re-calculated its standard errors using bootstrapp, model (4b) with the elapsed time, again, remarkably short ($\sim$84 seconds to fit a thousand models). While model (4) took roughly five seconds to be fitted, most of the computation time lies on calculating the support of the space of sufficient statistics. Once the the support of the sufficient statistics has been calculated, the optimization takes only a fraction of the time, which is why the bootstrap version of model (4) took about 0.09 seconds per repetition, and not 5.27 as the user may have expected. Details on the computational resources used for this section and the simulation studies are shown in \ref{sec:computing-details}.

\begin{figure}[p]
    \centering
    \includegraphics[width = .75\linewidth]{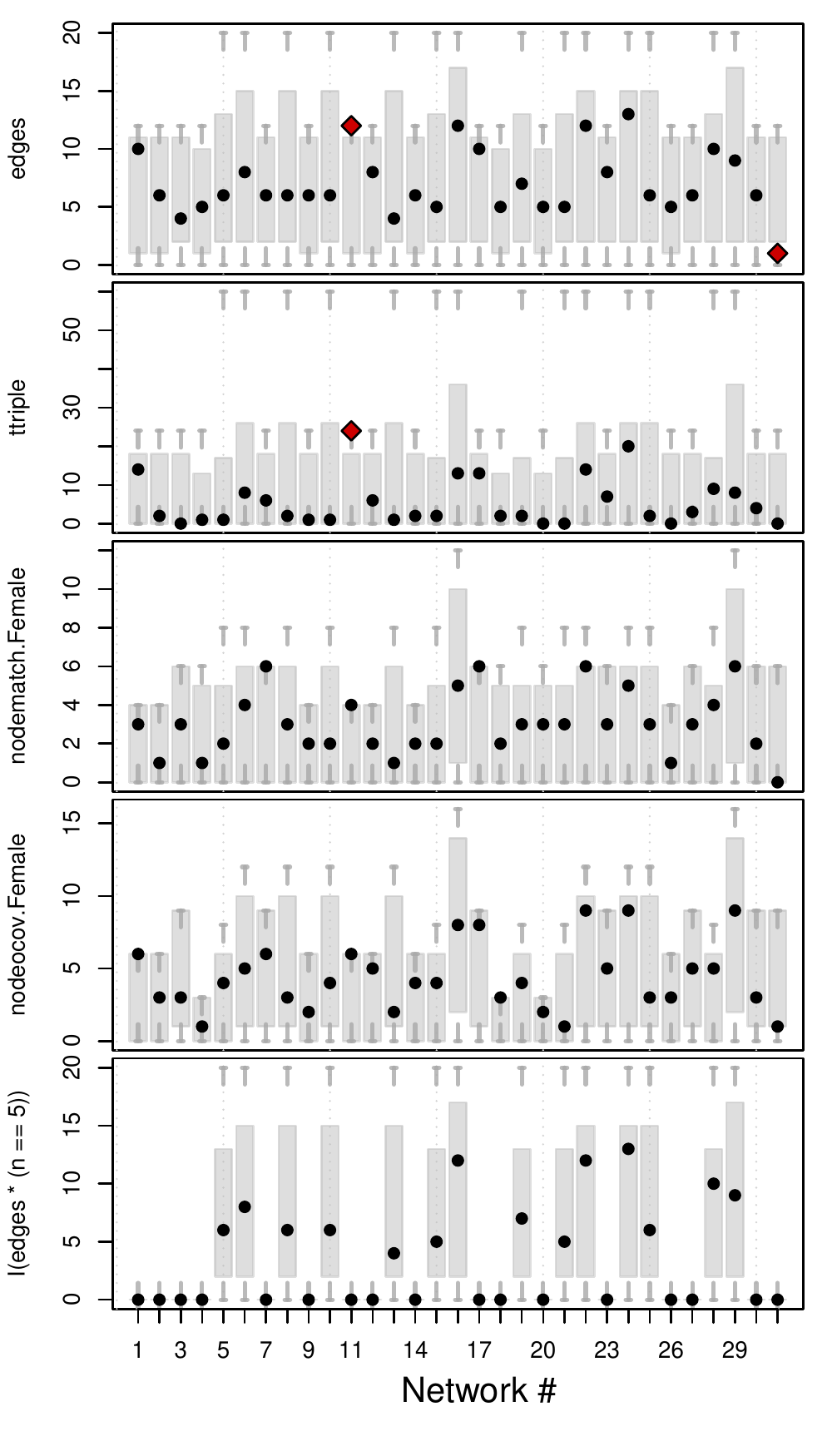}
    \caption{Distribution of the sufficient statistics under the ERGM specified by the parameters from model (4) in \autoref{tab:ci-ergm-full}. Each bar represents the exact 95\% confidence interval for the corresponding network+term combination, while the black dots show the location of the corresponding observed statistic. Red diamonds mark the observed statistics that fall out of the 95\% confidence interval. Of the 31 networks in the sample, it is only in two networks that the CIs don't cover the observed statistic, one that is fully connected and another that only has one tie.}
    \label{fig:ci-gof-full}
\end{figure}

As part of the \textit{ad hoc} diagnostics, \autoref{fig:ci-gof-full} shows the distribution of the sufficient statistic under the fitted model, 95\% exact confidence intervals (CI), versus the observed set of sufficient statistics. With the exception of two networks--one that is a full graph and another that only has one tie--the CIs generated by model (4) are able to cover all other network and term combinations.

We further discuss our results in the following section.

\section{Discussion}

In this paper we revisit and extend Exponential Family Random Graph Models (ERGM) for the case of small networks. Given the interest in testing hypotheses about small networks in the literature, but limited application of statistical models to small network data \cite[][and others]{Robins2007,Holland1981,Frank1986,Wasserman1996,Snijders2006}, we shed new light to ERGMs for small networks, which we call \ergmitos{}. An appealing feature of \ergmitos{} is that it allows direct use of the the full likelihood, with all that that entails, rather than costly approximations via MCMC methods. Overall, this approach provides a couple of important benefits for small network data: (1) it increases the chances of obtaining estimates in  the case that the observed sufficient statistics are near the boundary of its support (the convex-hull problem); as suggested by our simulation studies, (2) it has the potential to improve power and reduce type I error rates, compared to MC-MLE and the Robbins-Monro stochastic approximation; and (3) \ergmitos{} can be significantly faster to estimate (orders of magnitude faster), which in turn makes methods like bootstrap or other computationally intensive algorithms immediately available to be used with ERGMs, which to date has been unthinkable.

Another major benefit of using the full likelihood directly, when feasible, is that it gives researchers tremendous flexibility in terms of constructing and estimating new models. In terms of estimation, for example, the ability to easily calculate the likelihood allows researchers to make use of standard tools and techniques for ML estimation and MCMC estimation. This is important, because current techniques for intractable models (e.g., auxiliary variable MCMC, MC-MLE, and noise-contrastive estimation), while effective, are not necessarily straightforward to implement for non-experts. This places a high barrier to entry for researchers who would like to develop statistical models that go beyond the standard packages. 
As a simple example, take recent work on multilevel network models \cite{slaughter2016multilevel}. In that work, constructing a multilevel Bayesian model of a sample of networks, while conceptually straightforward, required the development of custom code and algorithms to implement. By contrast, having the full likelihood available in \textit{R} means that the same models studied in that paper (assuming small-N networks) can be constructed and estimated as easily as any other non-intractable model for which we can calculate the likelihood, using the full range of traditional tools and algorithms for ML and Bayesian estimation. This frees researchers from focusing only on models that are implementable in packages like \textit{statnet}, and allows greater freedom to think about ways that models for graphs can be modified and incorporated into other statistical models. In addition, being able to estimate gradients opens the possibility of estimating models using modern Bayesian algorithms like Hamiltonian Monte Carlo (HMC) and stochastic gradient langevin dynamics (SGLD), which may offer advantages in terms of speed or scalability, respectively.

The development and evaluation of \ergmitos{} in this simulation study also brings up topics for future work. One is the evaluation of model goodness-of-fit, and identifying statistics that are most important to evaluate with small networks, and that are reasonable to expect in a model that would suggest a ``good fit''. Because \ergmitos{} enable a rather simple way of conducting simulation studies (relative to traditional ERGMs), this will facilitate this work in future. Another topic to explore in future work is the value of \ergmitos{} for estimating ERGMs for very large networks, by drawing \textit{samples} of local network structures from a large graph. There is ongoing work extending ERGMs to very large networks \cite{STIVALA2016167,Stivala2020}, and \ergmitos{} could be a valuable approach for fitting these pooled models to a large sample (e.g., in the order of the thousands) of small local network structures drawn from a large network. Although this is an exciting extension to explore, this must proceed cautiously; while some matters as the ``projectivity problem'' \cite{shalizi2013} are solvable \cite{Krivitsky2011,Krivitsky2015,schweinberger2017note}, \textit{size-invariant} ERGMs (models in which the parameter estimates are \textit{scalable} across graph sizes), is an area of research very much under development.

In sum,  \ergmitos{} provide a promising extension to the ERGM framework for the analysis of small social networks. In addition to all the theoretical benefits that using exact likelihoods carry with it \cite{Handcock2003}, features related to computational efficiency and flexibility open the door to new \textit{and} old statistical tools that have been unreachable in the ERGM framework. Ultimately, as a fundamental building block of larger social systems, a richer understanding of the local social processes that give rise to the formation of networks in small social groups is key for our understanding of larger social structures that these constitute. 

\clearpage

\section{Acknowledgements}

% BLIND REVIEW

This material is based upon work support by, or in part by, the U.S. Army Research Laboratory and the U.S. Army Research Office under grant number W911NF-15-1-0577. The views, opinions, and/or findings contained in this paper are those of the authors and shall not be construed as an official Department of the Army position, policy, or decision, unless so designated by other documents

Computation for the work described in this paper was supported by the University of Southern California’s Center for High-Performance Computing (hpcc.usc.edu).

\nocite{R,butts2016,Wickham2016,Leifeld2013,VegaYon2019}
\printbibliography

\clearpage

\appendix

\section{MLE\label{appendix:mle}}

The estimation process of \ergmitos{} (as a pooled-data of small networks) is done entirely on R using the \textit{ergmito} R package. While a significant amount of the implementation of the methods described here was done using \texttt{Rcpp} \cite{Eddelbuettel2011}, a core component of the package is based on statnet's \textit{ergm} R package, and in particular, in the function \texttt{ergm.allstats} which does exhaustive enumeration of statistics in a compact way. In general, the estimation process for any list of networks is as follows:

\begin{enumerate}
    \item Analyze the model to be estimated: Extract the networks from the left-hand-side as specified in the ergm package, and calculate the exact statistics using the ergm.allstats function.
    \item With the full enumeration of statistics, build the joint likelihood function of the model in a compact form (i.e., using the weights instead of the full enumeration of the support of the model). This improves speed when it comes to evaluating the log-likelihood function.
    \item Because we are dealing with exact statistics, it is also possible to calculate the exact gradient function. We compute the gradient as follows:
    
    \begin{equation}
    \sum_{p}\nabla l_p(\theta) = \transpose{\sufstats{\adjmat_p, \Indepvar_p}} - \frac{\transpose{Q_p}\left(\transpose{W_p} \circ \exp{Q_p \theta}\right)}{\kappa_p}
    \end{equation}
    
    Where $\sufstats{\adjmat, \Indepvar}$ is a vector of observed sufficient statistics (usually called target statistics), $Q$ is a matrix of sufficient statistics, in particular, the isomorphic sufficient statistics associated with the model, and $W$ is a vector of frequency weights.
    
    These first three steps carry the most part of the computing time.
    
    \item Finally, the joint log-likelihood is maximized using the BFGS algorith implemented in the the optim function in the stats package.
    
\end{enumerate}

The final set of estimates is analyzed separately by another program included in the package. The next section describes the evaluation steps followed by \ergmito{}.

\section{\label{sec:evaluation-of-estimates}Evaluation of estimates}

After the optimization procedure finalizes, the \textit{ergmito} package performs a series of tests checking the quality of the estimates. In particular, we conduct the following evaluations after every call to the main optimization function:

\begin{enumerate}
	%\item Since the BFGS algorithm, as implemented in the \texttt{optim} function from the stats R package, only works with real numbers, we check whether the log-likelihood function increases with changes in the attained value in cases when the theoretical maxmima lies in $\pm\infty$. To do this, we increase the magnitude of each estimate by 1.5 and check the value of the log-likelihood function with that change. If an increase in that value is observed, we assume that the maxima for the given parameter equals $\mbox{sign}(\hat\theta_i)\times\infty$.
	%
	\item In the case that the observed sufficient statistics lied on the boundary of its support, the parameter estimate is set to be equal to the corresponding $\pm\infty$.
	
	\item If all parameters turn out to be $\pm\infty$ after this check, the function will send a warning message to the user and the function returns without computing the variance-covariance matrix. In general, the entries of the Hessian that involve a parameter estimate that diverge will be set to zero, which in turns results in zero variance-covariance for those entries when computing the Moore-Penrose generalized inverse. 
	
	\item If, on the other hand, a fraction of the parameters were switched to $\pm\infty$, the function recalculates the Hessian and the log-likelihood using the value $\mbox{sign}(\hat \theta_i)\times 10^{5}$. This is done instead of using $\infty$, because in most cases using infinite will result in the function being undefined. Again, the function will warn users about this issue. Nevertheless, this is mostly an implementation issue that we are already working on since the limiting values for the log-likelihood, gradient, and hessian are well defined in these cases (see \cite{Handcock2003}).
	
	\item In the case that the Hessian matrix is non-invertible (not positive-semi-definite [p.s.d.]), we use the Moore-Penrose generalized inverse algorithm as implemented in the R package MASS \cite{Venables2002}. For more on the interpretation of variance-covariance matrices when the Hessian is not p.s.d., see \cite{Gill2004}.
\end{enumerate}

The possible return codes are:

\begin{itemize}
\item[\textbf{00}] optim converged, no issues reported.
\item[\textbf{01}] optim converged, but the Hessian is not p.s.d.
\item[\textbf{10}] optim did not converged, but the estimates look OK.
\item[\textbf{11}] optim did not converged, and the Hessian is not p.s.d.
\item[\textbf{20}] A subset of the parameters estimates was replaced with +/-Inf.
\item[\textbf{21}] A subset of the parameters estimates was replaced with +/-Inf, and the Hessian matrix is not p.s.d.
\item[\textbf{30}] All parameters went to +/-Inf suggesting that the MLE may not exists.
\end{itemize}

\section{Computation details \label{sec:computing-details}}

All the simulations presented in this paper were executed in a large High-Performance-Computing cluster. In general, we use \textit{Slurm} \cite{Jette02slurm} over job arrays with 400 processors. Running the 20,000 simulations took about two hours on the cluster. 

In all other cases, i.e. not needing a large computing cluster, model fitting was done using a laptop computer with Ubuntu 18.04 LTS, 8GB of RAM, a quad-core processor Intel® Core i5-7200U CPU @ 2.50GHz, and using R version 3.6.3. The number of cores is relevant as the current implementation of the \textit{ergmito} R package uses \texttt{RcppArmadillo} R package \cite{Eddelbuettel2014} which can be compiled using OpenMP \cite{dagum1998openmp}, meaning that matrix algebra is multi-threaded.

\end{document}